\documentclass[letter,12pt]{article}
\usepackage{natbib}
\usepackage{url} 
\usepackage{adjustbox}
\usepackage{xcolor}
\usepackage{wrapfig}
\usepackage{multirow}
\usepackage[utf8]{inputenc}
\usepackage{graphicx}
\usepackage{dcolumn}
\usepackage{bbold}
\usepackage{bm}
\usepackage{svg}

\usepackage{setspace}
\usepackage{hyperref}

\usepackage{booktabs}

\usepackage{amsfonts,amsthm,amsmath,amssymb}
\usepackage{comment}
\usepackage{tabularx}
\usepackage{color}
\usepackage{wrapfig}
\usepackage{float}
\usepackage{algorithm} 
\usepackage{algpseudocode} 
\usepackage{appendix}
\usepackage{mathtools}
\usepackage{bbm}

\usepackage{caption}
\usepackage[flushleft]{threeparttable}

\newcommand{\ignore}[1]{}

\usepackage[english]{babel}

\newtheorem{theorem}{Theorem}
\newtheorem{lemma}{Lemma}

\newtheorem{example}{Example}
\newtheorem{assumption}{Assumption}

\DeclareMathOperator*{\argmax}{\arg\!\max} 

\newcommand{\MG}[1]{{\color{blue}MG: #1}}
\newcommand{\HL}[1]{{\color{red}#1}}

\usepackage{titlesec}

\titleformat{\section}                                     
  {\normalfont\Large\bfseries}
  {\thesection.}{1em}{}
\titleformat{\subsection}                                   
  {\normalfont\large\bfseries}
  {\thesubsection}{1em}{}

\newcommand{\blind}{0}

\newcommand{\beginappendix}{
\setcounter{table}{0}
\renewcommand{\thetable}{A\arabic{table}}
\setcounter{figure}{0}
\renewcommand{\thefigure}{A\arabic{figure}}
\setcounter{equation}{0}
\renewcommand{\theequation}{A\arabic{equation}}
\setcounter{lemma}{0}
\renewcommand{\thelemma}{n\arabic{lemma}}
\setcounter{section}{0}
\renewcommand{\thesection}{A\arabic{section}}
}

\usepackage[affil-it]{authblk}

\usepackage{xr}
\makeatletter

\newcommand*{\addFileDependency}[1]{
  \typeout{(#1)}
  \@addtofilelist{#1}
  \IfFileExists{#1}{}{\typeout{No file #1.}}
}
\makeatother

\newcommand*{\myexternaldocument}[1]{%
    \externaldocument{#1}%
    \addFileDependency{#1.tex}%
    \addFileDependency{#1.aux}%
}

\myexternaldocument{SKFCPD_supplement}

\begin{document}




\if0\blind
{
  \title{\vspace{-2.0cm}\bf  Sequential Kalman filter for fast online changepoint detection in longitudinal health records}

\author[1]{\vspace{-0.5cm}Hanmo Li}
\author[1]{Yuedong Wang}
\author[1]{Mengyang Gu\footnote{Corresponding authors: Yuedong Wang (\href{yuedong@pstat.ucsb.edu}{yuedong@pstat.ucsb.edu}) and Mengyang Gu (\href{mengyang@pstat.ucsb.edu}{mengyang@pstat.ucsb.edu}).}}

\affil[1]{Department of Statistics and Applied Probability, University of California, Santa Barbara}

\date{}
  \maketitle
} \fi

\if1\blind
{

  \title{\bf  Sequential Kalman filter for fast online changepoint detection in longitudinal health records}
  \date{}
  \author[]{}
  \maketitle
} \fi

\vspace{-.5in}
\begin{abstract}
This article introduces the sequential Kalman filter, a computationally scalable approach for online changepoint detection with temporally correlated data. The temporal correlation was 
not considered  in the Bayesian online changepoint detection approach due to the large computational cost. 
Motivated by detecting COVID-19 infections for dialysis patients from massive longitudinal health records with a large number of covariates, 
we develop a scalable approach to detect multiple changepoints from correlated data by sequentially stitching Kalman filters of subsequences to compute the joint distribution of the observations, which has linear computational complexity with respect to the number of observations between the last detected changepoint and the current observation at each time point, without approximating the likelihood function. Compared to other online changepoint detection methods, simulated experiments show that our approach is more precise in detecting single or multiple changes in mean, variance, or correlation for temporally correlated data. Furthermore, 
we propose a new way to integrate classification and changepoint detection approaches that improve the detection delay and accuracy for detecting COVID-19 infection compared to other alternatives. 



\end{abstract}

\noindent%
{\it Keywords:} Bayesian priors, dynamic linear models, Gaussian processes,  linear computational complexity, temporal correlation
\vfill



\newpage
\section{Introduction}
\label{sec:intro}

It is crucial to identify shifts in distribution proprieties from time series or longitudinal data, such as changes in mean, variance, and correlation, a process generally referred to as changepoint detection. 
Changepoint detection has become a widely utilized technique across various fields \cite{basseville1993detection}, including DNA copy number variants \citep{zhang2010detecting}, 
financial data \citep{fryz2014wild}, power systems \citep{chen2016quickest},  meteorology \citep{harris2022scalable} and cellular processes \citep{zhang2023modeling}, as a changepoint signals a deviation from the baseline data-generating process. 

In this work, we develop a computationally scalable and accurate approach to detect changepoints in time-dependent outcomes. 
Our aim is to detect whether a patient receiving dialysis treatment contracts severe acute respiratory syndrome coronavirus 2 (SARS-CoV-2) or COVID-19. 
The dialysis patient data are collected by Fresenius Kidney Care North America, which operates over 2,400 dialysis clinics in most US states and provides treatment for approximately one-third of the US dialysis patients. The dataset includes treatment and laboratory records for over 150,000 dialysis patients from January 2020 
to March 2022. 

We highlight that the longitudinal detection scenario considered herein is challenging as only around $0.4\%$ of the observations are in the COVID-19 infection periods formally defined in Section \ref{sec: COVID detection}, whereas other studies \citep{monaghan2021machine,Juntao2023Predicting}  consider the ``cross-sectional" observations, 
which matches one PCR test record to a few negative records, inducing a data set where $15\%-20\%$ of the records are COVID-19  positive, $30-50$ times higher than our setting which is closer to the real-world setting during pandemic. 
The low positive rates in the longitudinal setting make the detection of COVID-19 infections more challenging.
Our goal is to develop a new online changepoint detection method for identifying changes from longitudinal health records with a large number of measurements from lab tests, which are common in healthcare practice \citep{ni2020bayesian}. 
Various challenges exist for detecting COVID-19 infection, including large irregular missingness of time-dependent laboratory covariates of patients and temporal correlations in the probability sequences (see Figures \ref{fig:acf_pacf_covid} and \ref{fig:avg_acf_pacf_covid} of autocorrelation in the supplement material). 
To address these challenges, we propose an accurate and scalable changepoint detection algorithm 
that can integrate the results from state-of-the-art classification methods, such as XGBoost, and substantially improve the performance of classification methods. 
Our main interest lies in detecting changepoints as new data arrives sequentially, a key aspect distinguishing online from offline changepoint detection scenarios  
\citep{hinkley1970inference, fearnhead2006exact, killick2012optimal, fryz2014wild, matteson2014nonparametric, haynes2017computationally, hao2019asympotic}.
One popular framework of online change detection is the Bayesian online changepoint detection  \cite{fearnhead2007line,adams2007bayesian}, which was  shown to have high accuracy compared to other alternatives \citep{van2020evaluation}. 
However, one limitation of the Bayesian online changepoint detection is the assumption of mutual independence among observations, while correlations are common for temporal data. A few subsequent studies focus on detecting changepoints in the data with temporal correlations. For example, \cite{saatcci2010gaussian} utilizes a Gaussian process to model temporal correlation within the subsequences separated by changepoints. 
Although this approach reduces the computational complexity of detecting the change at each time step from $\mathcal{O}(n^4)$ to $\mathcal{O}(n^3)$  computational operations by using rank 1 updates \citep{scholkopf2002learning} for  $n$  observations, the computational complexity is still prohibitively large.  
\cite{fearnhead2011efficient} model the temporal correlations across segments using piecewise polynomial regression, assuming that correlations are Markov and observations within the same segments are independent.  \cite{romano2022detecting} model time series with autocorrelated noise and detects mean changes through dynamic programming recursion that maximizes the penalized likelihood. These approaches do not provide a flexible class of models of the temporal correlation between observations at each time point. 



Our main contributions are twofold. First, we propose efficient online changepoint detection algorithm, applicable for all dynamical linear models commonly used for modeling time sequences \citep{West1997,prado2010time}. 
The new algorithm is capable of sequentially detecting multiple changepoints with computational complexity $\mathcal{O}(n^{\prime})$ at each time, where $n^{\prime}$ is the number of observations between the last detected changepoint and the current observation, making it significantly more efficient than the Gaussian process changepoint detection  
algorithm \citep{saatcci2010gaussian}. 
We achieve this computational order by sequentially stitching Kalman filters of subsequences for computing likelihood and predictive distributions. 
This approach is generally applicable to all dynamic linear models with equally or unequally spaced time points. 
Second, when data contain a massive number of observations and high-dimensional covariates with a large proportion of missingness, it is challenging to apply any existing changepoint detection method or state space model directly. 
Our real application of detecting COVID-19 infection for dialysis patients is one such example, where a large number of lab covariates are missing as patients do not take all lab measurements in each of their visits.  To address this challenge, we propose an integrated approach. We first use supervised learning, such as XGBoost, to compute the posterior probability\footnote{{The posterior probabilities from the XGBoost model are calibrated by a sigmoid transformation to ensure they correspond well with the COVID-19 positive rate in the real-world dataset.}} of a time point being a changepoint for all patients. Conventional analysis often proceeds by choosing a threshold for the posterior probability to make detection decisions, which overlooks the changes in the longitudinal data that a changepoint detection algorithm could capture.
We apply our changepoint detection algorithm to each patient to detect changes in classification probabilities.  We found that the performance was dramatically improved compared to a supervised learning approach alone. The approach is general, as it's adaptable to any statistical machine learning method providing classification probabilities. 
Additionally, we provide \texttt{{SKFCPD}}, an R package for efficient implementation of our algorithm, to be released on CRAN along with the publication of this work.

This paper is structured as follows. Section \ref{sec: background} provides an overview of Bayesian online changepoint detection. In Sections \ref{sec: DLM}-\ref{sec: online update}, we introduce the sequential Kalman filter approach, an efficient online changepoint detection algorithm for temporally correlated data, and illustrate the computational advantage over direct computation in Section \ref{subsec: computation_complexity}. 
In Section \ref{sec: simulation}, we demonstrate the advantage of our proposed approach 
using simulated data with shifts in mean, variance, and correlation. 
Section \ref{sec: COVID detection} introduces the new approach that integrates classification methods with the new changepoint detection approach for  COVID-19 infection detection. 
Finally,  Section \ref{sec: conclusion}  introduces a few future directions. Proofs of lemmas, theorems, and additional numerical results are provided in the supplemental material.

\section{Online Changepoint Detection for Correlated Data}
\label{sec:fast_detection}

\subsection{Background: Bayesian online changepoint detection} 
\label{sec: background}

 \begin{figure}[]
\centering
\includegraphics[scale=.75]{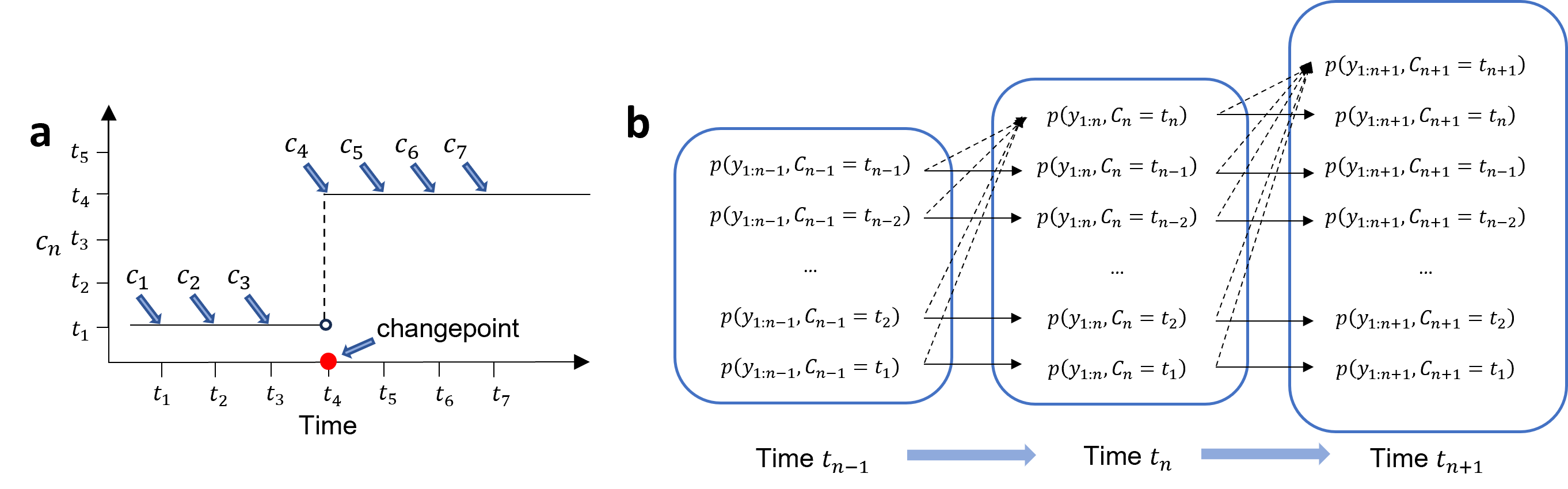}
\caption{Panel a: value of the state $C_n$ that shows the most recent changepoint before or at the time $t_n$ for $n=1,\dots, 7$. The time point $t_4$, marked by a red dot, is the only changepoint before $t_7$. Panel b: the recursive process of computing the joint distribution $p(\mathbf{y}_{1:n}, C_n)$ from time $t_{n-1}$ to $t_{n+1}$ based on the Equation (\ref{equ: joint_dist_BOCPD_3}). The black arrow means the latter probability can be sequentially computed from the former one. 
}
\label{fig: iterative updating}
\end{figure}
 
Let us consider the time series 
$\mathbf{y}_{1:n} = \left(y_{1}, \dots, y_{n} \right)^T \in \mathbb{R}^n$ for   time points $\{t_1, \dots, t_n\}$, such that $t_j<t_i$ for any $1\leq j<i\leq n$. 
We assume  time segments separated by any two changepoints are independent of each other, whereas data within each segment can be temporally correlated. Each segment can have distinct distributions characterized by different mean, variance, or correlation parameters.

We define $C_n$ as the most recent changepoint at or before the current time point $t_n$. 
{For instance, if $C_n=t_4$, it indicates that $t_4$ is the only changepoint in the time period $[t_4, t_n]$. As shown in Figure \ref{fig: iterative updating}a, where time $t_4$ is the only changepoint before time $t_7$, $C_n$ shifts from $t_1$ to $t_4$ at time $t_4$.}  We define run length, $r_n$, as the length of the time interval from the most recent changepoint to the current time point, calculated as $r_n=n-C_n+1$.

 The objective of online changepoint detection is to sequentially estimate the changepoint $C_n$ upon receiving a new observation at the current time $t_n$. A popular online changepoint detection framework is the BOCPD method \citep{fearnhead2007line,adams2007bayesian}, which has a few assumptions. 
\begin{assumption}
{The segments partitioned by changepoints are mutually independent.} 
\label{assumption:independence_segments}
\end{assumption}

\begin{assumption}
{The state on the current time point $C_n$, conditioning on the state of the previous time point $C_{n-1}$, is independent of the observations of $\mathbf y_{1:(n-1)}$.} 
\label{assumption:run_length}
\end{assumption}

 {Based on the second assumption, given the previous state $C_{n-1}=t_j$ for $1 \leq j \leq n-1$, $C_{n}$ can either be $t_n$ if $t_n$ is a changepoint, or remain as $t_j$ if $t_n$ is not a changepoint. Thus, $C_n$ is restricted to either $t_j$ or $t_n$. 
Following the  BOCPD framework, we define the prior distribution of the conditional distribution of the most recent changepoint as $p(C_n = t_i \mid C_{n-1}=t_j)$, where it takes the value of  $1 - H(t_i)$ if $i=j$, $H(t_i)$ if $i=n$ and is zero in all other cases. $H(\cdot)$ is the hazard function, measuring the probability that a changepoint occurs at any time point.

{In BOCPD, the hazard function is often defined as $H(t_i) = \frac{1}{\lambda_i}$, where $\frac{1}{\lambda_i}$ represents the prior probability of time $t_i$ being a changepoint, typically held fixed in practice. 
For applications such as detecting COVID-19 infection, a time-dependent hazard function can be used to integrate local infection information.}

We allow the observations to be mutually dependent within each segment of the changepoints, which relaxes the additional assumption of independence between each observation within one segment in \cite{adams2007bayesian}. This modification offers a more realistic modeling for time series data, where observations are often correlated. Furthermore, Assumption 2  means information from the previous observations $\mathbf y_{1:(n-1)}$ is contained in  $C_{n-1}$, the latent state indicating whether the previous time point is a changepoint.  
Based on Assumptions 1 and 2  
from BOCPD,  we compute  the joint distribution of the state $C_n=t_i$ and the observations $\mathbf{y}_{1:n}$  by integrating out the previous state $C_{n-1} = t_j$, 
\begin{align}
         &p\left(\mathbf{y}_{1:n}, C_n = t_i \right) \nonumber\\
         = &\underbrace{p\left(y_{n} \mid \mathbf{y}_{i:(n-1)}, C_{n}=t_i\right)}_{\text{predictive distribution}} \sum_{j=1}^{n-1} \underbrace{p\left(C_n=t_i \mid  C_{n-1}=t_j\right)}_{\text{hazard}}   p\left(\mathbf{y}_{1:(n-1)},C_{n-1}=t_j\right) \nonumber\\
         = &\begin{cases} 
      p\left(y_n \mid \mathbf{y}_{i:(n-1)} , C_{n} = t_i\right)  \left(1- H(t_i)\right) p\left(\mathbf{y}_{1:(n-1)}, C_{n-1}=t_i\right), & i<n, \\
       p\left(y_n \mid C_n=t_n\right)  H(t_n)\sum_{j=1}^{n-1}  p\left(\mathbf{y}_{1:(n-1)}, C_{n-1}=t_j\right), & i = n, \label{equ: joint_dist_BOCPD_3}
   \end{cases}
\end{align}
where the derivation  is given  in Section \ref{supsec: proof_of_equation_3} of the supplementary material.

After obtaining the joint probability $p(\mathbf{y}_{1:n}, C_n=t_i)$ for $i=1,\dots,n$,  one can estimate the state $\hat{C}_n$ by calculating the maximum \textit{a posteriori} (MAP) estimate of the joint distribution \citep{fearnhead2007line}, 
\begin{equation}
\begin{aligned}
   \hat{C}_n 
   & = \argmax_{t_1\leq t_i\leq t_n} p(\mathbf{y}_{1:n}, C_n=t_i).
\end{aligned}
\end{equation}

The probability of $p(\mathbf{y}_{1:n}, C_n=t_i)$ needs to be computed for all possible time points $t_i$, for $i=1,...,n$, to obtain the MAP of the changepoint upon receiving new data at time $t_n$.    Figure \ref{fig: iterative updating}b shows 
 the recursive computational process for the joint probability $p(\mathbf{y}_{1:n}, C_n=t_i)$ in Equation (\ref{equ: joint_dist_BOCPD_3}). At time $t_n$, we can recursively compute the probability $p(\mathbf{y}_{1:n}, C_n=t_i)$  from the previous step $p(\mathbf{y}_{1:(n-1)}, C_{n-1}=t_i)$, where $i<n$, as indicated by the solid black arrows.
  Furthermore, 
  the probability $p(\mathbf{y}_{1:n}, C_n=t_n)$ can be computed through probabilities of changepoints occurring at previous time points $\{p(\mathbf{y}_{1:(n-1)}, C_{n-1}=t_j)\}_{j=1}^{n-1}$ and the marginal distribution of the current time point being a changepoint $p(y_n\mid C_n=t_n)$, as shown by the black dashed arrows in Figure \ref{fig: iterative updating}b.

{
  By considering different combinations of changepoints in the joint distribution $p(\mathbf{y}_{1:n}, C_n)$, the recursive formula in Equation (\ref{equ: joint_dist_BOCPD_3})  enables the algorithm to sequentially detect multiple changepoints. 
 For problems with multiple changepoints, we may exclude the observations prior to the most recently detected changepoint to further reduce computational complexity. Specifically,  instead of summing over all time indices $j$ from $1$ to $n-1$ to compute the joint distribution in Equation (\ref{equ: joint_dist_BOCPD_3}), we can truncate the summation to the range from $t_{\hat j}$ to $n-1$, where $t_{\hat j}=\hat{C}_{n-1}$ denotes the most recently detected changepoint at or before time $t_{n-1}$. {Note that the estimated most recent changepoint at  $(n-1)$th time point $t_{\hat j}=\hat{C}_{n-1}$ and $t_j=C_{n-1}$ defined in Equation (\ref{equ: joint_dist_BOCPD_3}) can be different. 
 } 
 {This approach is more efficient, as two subsequences separated by a changepoint are mutually independent and their distributions can contain distinct parameters. 
Both simulated and real data studies validate that this approach effectively reduces the computational cost for detecting multiple changepoints without compromising accuracy.
} 


The computation of the joint probability $p(\mathbf{y}_{1:n}, C_n = t_i)$ in Equation (\ref{equ: joint_dist_BOCPD_3}) demands an efficiently evaluation of prediction probabilities $\{p(y_{n}\mid \mathbf{y}_{i:(n-1)}, C_{n}=t_i)\}_{i=1}^{n-1}$.
To scalably compute the joint distribution, \cite{adams2007bayesian} assumed the observations are independent and identically distributed (i.i.d.) random variables with the exponential family of distributions.
However, the i.i.d. assumption of observations may not hold for many real-world datasets. To address temporal correlations in time sequences, \cite{fearnhead2007line} proposed a method that utilizes the particle filter \citep{doucet2000sequential}  
to approximate the predictive distributions, which may compromise accuracy due to the approximation of the likelihood and the choice of the inducing inputs. 
To overcome these challenges, we propose a new approach for online changepoint detection applicable for dynamic linear models to model temporally correlated data, 
which efficiently computes the predictive distributions without approximating the likelihood function. 

\subsection{Dynamic Linear Models for Online Changepoint Detection}
\label{sec: DLM}

Gaussian processes (GPs) have been used to model temporally correlated measurements for online changepoint detection  \citep{saatcci2010gaussian}. 
By Assumption \ref{assumption:independence_segments}, the GP model can have different parameters across segments. The marginal distribution of the $(m+1)$-th segment follows a multivariate normal distribution  $\left( \left(y(t_{\tau_{m}}), \dots, y(t_{\tau_{m+1}-1})\right)^T \mid \mu_m, \sigma_m^2, \gamma_{m}, \sigma_{0,m}^2 \right) \sim \mathcal{MN}(\mu_m, \sigma_m^2 
\mathbf{R}_{\tau_{m+1} - \tau_{m}}+ \sigma_{0, m}^2 \mathbf{I}_{\tau_{m+1} - \tau_{m}})$, where $\mathbf{R}_{\tau_{m+1} - \tau_{m}}$ is a  $(\tau_{m+1} - \tau_{m})\times(\tau_{m+1} - \tau_{m})$ correlation matrix having parameter $\gamma_m$, with $\tau_m$ being the time index of the $m$th changepoint,  for $m=1,\dots, M-1$, and when $m=0$, we let $\tau_0=1$.  For simplicity, we focus on the time range from $t_i$ to $t_n$, where $t_i$ is larger than the previously detected changepoint $\hat C_{n-1}$. 
The total number of observations within this segment is denoted by $n^{\prime}=n-i+1$. The subscript $m$ in the  parameters $(\mu_m, \sigma_m^2, \gamma_m,\sigma^2_{0,m})$ will be dropped to  simplify the notations.

 Directly calculating the likelihood and predictive distributions by a GP, however, can be computationally expensive, as it requires computing inversion of the covariance matrix $\textbf{R}_{n^{\prime}}$, which takes $\mathcal{O}({n^{\prime}}^3)$ operations. The predictive distribution of the online changepoint detection needs to be calculated numerous times, which further exacerbates the computational challenge. 
Here we model the temporally dependent observations by dynamic linear models (DLMs)  \citep{West1997,stroud2001dynamic,durbin2012time},  a large class of models for scalable computation. 

For simplicity, we denote $y_{i+k-1}=y(t_{i+k-1})$, the real-valued observation at time $t_{i+k-1}$,  which does not need to be equally spaced, for $k=1,\dots,n^{\prime}$. We consider a DLM below, 
\begin{equation}
\begin{aligned}
&y_{i+k-1}=\mu + \textbf{F}_k \bm \theta_k+\epsilon_k, \quad \epsilon_k \sim \mathcal{N}\left(0, \sigma_0^2\right), \\
&\bm \theta_k=\textbf{G}_k \bm \theta_{k-1}+\textbf{w}_k, \quad \textbf{w}_k \sim \mathcal{MN}\left(0, \textbf{W}_k\right),
\end{aligned}
\label{equ: DLM}
\end{equation}
where $\mu$ is the mean parameter, $\bm \theta_k$ is a $q$-dimensional latent state process with {the initial state $\bm \theta_{0} \sim \mathcal{N}(\mathbf 0, \mathbf{B}_{0})$},  $\textbf{F}_k$ is a $1 \times q$ vector, $\mathbf{B}_{0}$, and $\textbf{G}_k$ and  $\textbf{w}_k$ are $q\times q$ matrices.

As an example, a GP with a Mat{\'e}rn covariance function that contains half-integer roughness parameters \citep{handcock1993bayesian,stein1999interpolation} can be written as a DLM \citep{whittle1954stationary,hartikainen2010kalman}. For instance,  the  Mat{\'e}rn covariance function with a roughness parameter being 0.5 follows 
\begin{equation}   
   \sigma^2 c(t, t^{\prime}) = \sigma^2  \exp{\left(-\frac{|d|}{\gamma}\right)},
    \label{equ: exp_kernel}
\end{equation}
where $d=t-t^{\prime}$ and $\gamma$ is the range parameter, for any $t$ and $t^{\prime}$. 
The GP  with covariance in (\ref{equ: exp_kernel}) is equivalent to a DLM in Equation (\ref{equ: DLM}) with   ${F}_k = 1$, ${G}_k=\rho_k$, ${W}_k = \sigma^2 \left(1 - \rho^2_k\right)$,   $\rho_k=\exp{\left(-\frac{|t_{i+k-1}-t_{i+k-2}|}{\gamma}\right)}$, 
 and ${B}_0 = \sigma^2$.  

The Mat{\'e}rn with roughness parameter being 2.5,  as another example,  follows  
\begin{equation}
 \sigma^2 c(t, t') = \sigma^2 \left(1+\frac{\sqrt{5} |d|}{\gamma}+\frac{5 d^2}{3 \gamma^2}\right) \exp \left(-\frac{\sqrt{5} |d|}{\gamma}\right).
    \label{equ: Matern_kernel}
\end{equation} 
The equivalent representation of a DLM  is discussed in Section \ref{sec: connection_DLM_and_Matern} of the supplementary material. 
We will use GPs with the two covariance functions in Equations (\ref{equ: exp_kernel}) and (\ref{equ: Matern_kernel})  for illustrative purposes, but our approach is generally applicable to all DLMs, which includes a much larger class of processes. 

From Equation (\ref{equ: joint_dist_BOCPD_3}), to evaluate the joint distribution $p\left(\mathbf{y}_{1:n}, C_n = t_i \right)$ for the last segment $(t_i, \dots, t_n)$ where $C_n=t_i$ is the most recent changepoint prior to $t_n$, we need to efficiently compute the predictive distribution $p\left(y_n \mid \mathbf{y}_{i:(n-1)}\right)$. 
For the computational reason, we define the noise variance to signal variance ratio  $\eta = \frac{\sigma_0^2}{\sigma^2}$, and the covariance matrix of observations $\mathbf{y}_{i:n}$ is given by  
\begin{equation}
   \sigma^2 \mathbf{K}_{n^{\prime}} = \sigma^2(\mathbf{R}_{n^{\prime}} + \eta \mathbf{I}_{n^{\prime}}).
    \label{equ: covariance_of_y_1_n}
\end{equation}
After this transformation, the parameter set is given by $\bm \Theta = \left\{\mu, \sigma^2, \gamma, \eta\right\}$. In the following, we present the direct computation of the predictive distribution $p\left(y_n \mid \mathbf{y}_{i:(n-1)}\right)$. 

Assuming the objective prior for the mean and variance parameter $p(\mu, \sigma^2) \propto \frac{1}{\sigma^2}$, the predictive distribution $p\left(y_{n} \mid \mathbf{y}_{i:(n-1)}, \gamma, \eta\right)$, after integrating out $(\mu, \sigma^2)$, follows,
\begin{align}
        &p\left(y_{n} \mid \mathbf{y}_{i:(n-1)}, \gamma, \eta\right)
         = \frac{p\left(\mathbf{y}_{i:n} \mid \gamma, \eta\right)}{p\left(\mathbf{y}_{i:(n-1)} \mid \gamma, \eta\right)} \nonumber \\
        & \propto  \begin{cases} 
      \frac{\Gamma(\frac{n^{\prime}-1}{2})}{\Gamma(\frac{n^{\prime}-2}{2})}  \left(\frac{|\mathbf{K}_{n^{\prime}}|}{|\mathbf{K}_{n^{\prime}-1}|}\right)^{-1/2} 
        \left(\frac{\mathbf{1}_{n^{\prime}}^T \mathbf{K}_{n^{\prime}}^{-1} \mathbf{1}_{n^{\prime}}}{\mathbf{1}_{n^{\prime}-1}^T \mathbf{K}_{n^{\prime}-1}^{-1}\mathbf{1}_{n^{\prime}-1}}\right)^{-1/2} \exp \left(-S_{n^{\prime}}^2 \right), & i < n-1  \\
       \left(\frac{|\mathbf{K}_{n^{\prime}}|}{|\mathbf{K}_{n^{\prime}-1}|}\right)^{-1/2} 
        \left(\frac{\mathbf{1}_{n^{\prime}}^T \mathbf{K}_{n^{\prime}}^{-1} \mathbf{1}_{n^{\prime}}}{\mathbf{1}_{n^{\prime}-1}^T \mathbf{K}_{n^{\prime}-1}^{-1}\mathbf{1}_{n^{\prime}-1}}\right)^{-1/2} \left(\mathbf{y}_{(n-1):n}^T \mathbf{M}_{n^\prime} \mathbf{y}_{(n-1):n}\right)^{-1/2}, & i = n-1 
   \end{cases} \label{equ: pred dist model 2}
\end{align}
where $   S_{n^{\prime}}^2 = \left(\frac{n^{\prime}-1}{2}\right) \log \left(\mathbf{y}_{i:n}^T \mathbf{M}_{n^{\prime}} \mathbf{y}_{i:n}\right) - \left(\frac{n^{\prime}-2}{2}\right) \log\left(\mathbf{y}_{i:(n-1)}^T \mathbf{M}_{n^{\prime}-1} \mathbf{y}_{i:(n-1)}\right)$ and 
 $  \mathbf{M}_{n^{\prime}}  = \mathbf{K}_{n^{\prime}}^{-1} 
  - \mathbf{K}_{n^{\prime}}^{-1} \mathbf{1}_{n^{\prime}} \left(\mathbf{1}_{n^{\prime}}^T \mathbf{K}_{n^{\prime}}^{-1} \mathbf{1}_{n^{\prime}}\right)^{-1} \mathbf{1}_{n^{\prime}}^T \mathbf{K}_{n^{\prime}}^{-1}.$ {Note that when $i = n-1$, there is not enough information to simultaneously integrate out $\mu$ and $\sigma^2$ for $p(y_n \mid y_{n-1})$. We develop a new procedure for this problem. When $i = n-1$, we first integrate out $\mu$ in the joint distribution using the prior probability $\pi(\mu) \propto 1$. Then, we integrate out $\sigma^2$ in the predictive distribution $p(y_n \mid y_{n-1})$ using the prior probability $\pi(\sigma^2) \propto \frac{1}{\sigma^2}$. Through simulation and real data analysis, we found that this  procedure at $i = n-1$ provides a stable evaluation of the predictive distribution and avoids mistakenly detected changepoints.}
The derivation of Equation (\ref{equ: pred dist model 2})  is given in Section \ref{sec: Proof equ: pred dist model 2} in the supplementary material. 

{Directly applying Equation (\ref{equ: pred dist model 2}) to compute the predictive distribution requires $\mathcal{O}\left(n^{\prime 3}\right)$ computational operations, due to matrix inversion and determinant calculation. This makes the computation impractical as the predictive distribution must be computed for all previous time points. }
 In the following section, we develop the sequential Kalman filter to improve the computational efficiency of the Equation (\ref{equ: pred dist model 2}) without any approximation.

\subsection{Sequential Kalman filter for Fast Changepoint Detection}
\label{sec: online update}

 In this section, we introduce a fast algorithm, called sequential Kalman filter (SKF) to reduce the complexity of computing    $p\left(y_{n} \mid \mathbf{y}_{i:(n-1)}, \gamma, \eta\right)$ from $\mathcal{O}(n^{\prime 3})$ to $\mathcal{O}(1)$ with $n'=n-i+1$, for each $i=1,\dots, n-1$. 
First, we discuss the Cholesky decomposition to draw the connection between the predictive distribution and the Kalman filter (KF).
Denote the Cholesky decomposition of the covariance matrix as $\mathbf{K}_{n^{\prime}} = \mathbf{L}_{n^{\prime}} \mathbf{L}_{n^{\prime}}^T$, where $\mathbf{L}_{n^{\prime}}$ is an $n^{\prime}\times n^{\prime}$ lower triangular matrix. Consequently, the inverse covariance matrix can be decomposed as $\mathbf{K}_{n^{\prime}}^{-1}= \mathbf{U}_{n^{\prime}}^T \mathbf{U}_{n^{\prime}}$, 
where $\mathbf{U}_{n^{\prime}} = \mathbf{L}_{n^{\prime}}^{-1}$.
As computing Cholesky decomposition takes $\mathcal O(n'^3)$ operations, we extend the KF in Lemma \ref{lemma:a_b_k} and Theorem \ref{thm: KF update} to compute two $n^{\prime}$-vectors  $\mathbf{u}_{n^{\prime}}  = \mathbf{U}_{n^{\prime}}\mathbf{1}_{n^{\prime}}$ and 
$\mathbf{v}_{i, n^{\prime}}  = \mathbf{U}_{n^{\prime}}\mathbf{y}_{i:n}$, where $\mathbf{u}_{n^{\prime}} = \left( u_1, \dots, u_{n^{\prime}} \right)^T$ and $\mathbf{v}_{i, n^{\prime}} = \left( v_{i, 1}, \dots, v_{i, n^{\prime}} \right)^T$ are both $n^{\prime}$-vectors. 
 The proofs for Lemma \ref{lemma:a_b_k} and Theorem \ref{thm: KF update} are given in  Sections  \ref{sec: proof_of_Lemma_1} and \ref{sec: proof_thm_1} of the supplementary material, respectively. 

\begin{lemma}
For $k = 1, \dots, n^{\prime}$, the $k$th element of $\mathbf{u}_{n^{\prime}}=\mathbf{U}_{n^{\prime}}\mathbf{1}_{n^{\prime}}$ and $\mathbf{v}_{i, n^{\prime}}=\mathbf{U}_{n^{\prime}}\mathbf{y}_{i:n}$ can be sequentially computed as follows  
    \begin{align}
        u_{k} & =  \frac{1 - f_k^u}{\sqrt{Q_k^u}}, \label{equ:a_k}\\
        v_{i, k} & = \frac{y_{i+k-1} - f_{i, k}^v}{\sqrt{Q_{i, k}^v}}, \label{equ:b_k}
    \end{align}
where for $k\ge2$, we have
    $f_k^u  = \mathbb{E}_{\mathbf{Y}_{1:k}}[Y_{k} \mid \mathbf{Y}_{1:(k-1)} = \mathbf{1}_{k-1}, \gamma, \eta] = g_k^u( f_{k-1}^u, Q_{k-1}^u),$ 
    $Q_k^u  = \mathbb{V}_{\mathbf{Y}_{1:k}}[Y_{k}  \mid \mathbf{Y}_{1:(k-1)} = \mathbf{1}_{k-1}, \gamma, \eta] = h_k^u(Q_{k-1}^u), $
    $f_{i,k}^v= \mathbb{E}_{\mathbf{Y}_{1:k}}[Y_{k}  \mid \mathbf{Y}_{1:(k-1)} = \mathbf{y}_{i:(i+k-2)}, \gamma, \eta] = g_{i, k}^v\left(f_{i, k-1}^v, Q_{i, k-1}^v\right), $ and
    $Q_{i, k}^v  = \mathbb{V}_{\mathbf{Y}_{1:k}}[Y_{k} \mid \mathbf{Y}_{1:(k-1)} = \mathbf{y}_{i:(i+k-2)}, \gamma, \eta] = h_{i, k}^v(Q_{i, k-1}^v),$
with   $\mathbf{Y}_{1:n^{\prime}}$ denotes 
a random output vector in a DLM with covariance $\mathbf{K}_{n'}$.
The  functions $g_k^u(\cdot)$, $h_k^u(\cdot)$, $g_{i, k}^v(\cdot)$, and $h_{i, k}^v(\cdot)$ are given in Equations (\ref{equ: f_k_u})-(\ref{equ: q_k_v}) of the supplementary material.

\label{lemma:a_b_k}
\end{lemma}

In Lemma \ref{lemma:a_b_k}, the KF is iteratively applied for  computing the parameters $f_k^u$, $Q_k^u$, $f_{i, k}^v$, and $Q_{i, k}^v$ from the  parameters at the previous time point $f_{k-1}^u$, $Q_{k-1}^u$, $f_{i, k-1}^v$, and $Q_{i, k-1}^v$. 
Once we obtain these parameters, $u_k$ and $v_{i, k}$ can be computed with  $\mathcal{O}(1)$ operations for each $k=1,\dots,n^{\prime}$ by using Equations (\ref{equ:a_k}) and (\ref{equ:b_k}), respectively.  The derivation of Lemma \ref{lemma:a_b_k} is provided in Section  \ref{sec: proof_of_Lemma_1} in the supplementary material.

\begin{theorem}

 After obtaining each term of $\mathbf{u}_{n^{\prime}}$ and $\mathbf{v}_{i, n^{\prime}}$ from Equations (\ref{equ:a_k}) and (\ref{equ:b_k}), the predictive distribution in Equation (\ref{equ: pred dist model 2}) can be computed below 
\begin{equation}
    p(y_{n} \mid \mathbf{y}_{i:(n-1)}, \gamma,\eta)
     \propto \begin{cases}
         \frac{\Gamma(\frac{n^{\prime}-1}{2})}{\Gamma(\frac{n^{\prime}-2}{2})} \left(Q_{n^{\prime}}^u\right)^{-\frac{1}{2}}
     \left(\frac{\mathbf{u}_{n^{\prime}}^T \mathbf{u}_{n^{\prime}}}{\mathbf{u}_{n^{\prime}-1}^T \mathbf{u}_{n^{\prime}-1}}\right)^{-1/2} \exp\left(-S_{n^{\prime}}^2 \right), & i<n-1\\
     \left(Q_{n^{\prime}}^u\right)^{-\frac{1}{2}}
     \left(\frac{\mathbf{u}_{n^{\prime}}^T \mathbf{u}_{n^{\prime}}}{\mathbf{u}_{n^{\prime}-1}^T \mathbf{u}_{n^{\prime}-1}}\right)^{-1/2} \left(\mathbf{y}_{i:n}^T \mathbf{M}_{n^{\prime}} \mathbf{y}_{i:n} \right)^{-1/2}, & i=n-1\\
     \end{cases}
     \label{equ: pred_distribution_kf_objective_prior}
\end{equation}
where $ S^{2}_{n^{\prime}} = \left(\frac{n^{\prime}-1}{2}\right) \log\left(\mathbf{y}_{i:n}^T \mathbf{M}_{n^{\prime}} \mathbf{y}_{i:n}\right) - \left(\frac{n^{\prime}-2}{2}\right) \log\left(\mathbf{y}_{i:(n-1)}^T \mathbf{M}_{n^{\prime}-1} \mathbf{y}_{i:(n-1)}\right)
$ and $    \mathbf{y}_{i:n}^T \mathbf{M}_{n^{\prime}} \mathbf{y}_{i:n} 
    = \mathbf{v}_{i, n^{\prime}}^T \mathbf{v}_{i, n^{\prime}} - (\mathbf{u}_{n^{\prime}}^T \mathbf{u}_{n^{\prime}})^{-1} (\mathbf{v}_{i, n^{\prime}}^T\mathbf{u}_{n^{\prime}})^2
$. 

\label{thm: KF update}
\end{theorem}

When a new observation $y_{n}$ is available at time $t_n$, we apply  Lemma \ref{lemma:a_b_k} to update the variables $\mathbf{u}_{n^{\prime}}$ and $\mathbf{v}_{i, n^{\prime}}$ with  $\mathcal{O}(1)$ operations, and compute the predictive distribution $p(y_{n} \mid \mathbf{y}_{i:(n-1)}, \gamma, \eta)$ based on Equation (\ref{equ: pred_distribution_kf_objective_prior}), which is significantly faster than directly computing the inversion of the covariance matrix in Equation (\ref{equ: pred dist model 2}). 

As the estimation of range and nugget parameter $(\gamma, \eta)$ typically does not have closed-form expressions,
we maximize the likelihood function over a set of training samples from an initial or control period containing no changepoint:
\begin{equation}
(\hat \gamma, \hat \eta) 
= \argmax_{(\gamma, \eta)} p( \mathbf{y}_{\mathcal S_{tr}} \mid \gamma, \eta),
\label{equ:mle}
\end{equation}
where $\mathcal S_{tr}$ is an index set of $n_{tr}$  time indices in the training time period.
We employ the KF to compute the likelihood function, which only requires $\mathcal O(n_{tr})$ operations \citep{gu2022scalable}. 
We plug the estimated range and nugget parameters into the SKF algorithm for online changepoint detection, effectively capturing the temporal correlations in the data. It important to note that the mean and variance parameters in the SKF algorithm are integrated out based on all available observations, which enables the algorithm to incorporate the latest information for online changepoint detection. To avoid large computational costs, the range parameters and nugget parameters were estimated using training sequences, similar to the GPCPD approach. In Section \ref{sec: simulation}, we empirically show that SKF can accurately detect mean, variance, and correlation changes.  




\begin{algorithm}[t]
	\caption{{Sequential Kalman Filter algorithm}  for fast changepoint detection} 
	\hspace*{\algorithmicindent} \textbf{Input:} New observation $y_{n}$, previously estimated changepoint time index $\hat{j}$, previous  parameters $f^u_{n-1}$,$Q^u_{n-1}$, $f^v_{i, n-1}$ and $Q^v_{i, n-1}$ for $\hat{j} \leq i \leq n-1$ 
 defined in Equation (\ref{equ: KF_param_update}), the  joint distribution $p(\mathbf{y}_{1:(n-1)},C_{n-1} \mid \hat \gamma,\hat \eta)$,  estimated nugget and range parameters $(\hat \gamma,\hat \eta)$
	\\
    \hspace*{\algorithmicindent} \textbf{Output:} The estimated most recent changepoint $\hat{C}_n$, current  parameters $f^u_{n}$,$Q^u_{n}$, $f^v_{i, n}$ and $Q^v_{i, n}$ for $\hat{j} \leq i \leq n-1$ and the joint distribution $p(\mathbf{y}_{1:n},C_{n} \mid \hat \gamma,\hat \eta)$
	\begin{enumerate}
		\item  \textbf{Update parameters through Kalman filter}
		
		We iteratively compute  parameters $\left( f^u_{n}, Q^u_{n}, f^v_{i, n}, Q^v_{i, n}\right)$ from $\left( f^u_{n-1}, Q^u_{n-1},\right.$ $\left. f^v_{i, n-1}, Q^v_{i, n-1}\right)$ for $\hat{j} \leq i \leq n-1$ by Lemma \ref{lemma:a_b_k}.

	    \item  \textbf{Compute  predictive distributions} 
	    
	   We sequentially compute the predictive distribution $p(y_{n} \mid \mathbf{y}_{i:(n-1)}, \hat \gamma,\hat \eta) $ by parameters $f^u_{n}$,$Q^u_{n}$, $f^v_{i, n}$ and $Q^v_{i, n}$ based on Equation (\ref{equ: pred_distribution_kf_objective_prior}), for $\hat{j} \leq i \leq n-1$.
	    
	    \item \textbf{Update  joint distributions} 
     
	    When $t_n$ is not a changepoint, we have $C_n < t_n$. 
     For $\hat{j} \leq i \leq n-1$, 
	    \begin{equation}
     \begin{aligned}
         &p(\mathbf{y}_{1:n}, C_n=t_i \mid \hat \gamma,\hat \eta)= p(y_{n} \mid \mathbf{y}_{i:(n-1)}, \hat \gamma,\hat \eta) \left(1- H(t_i)\right) p(\mathbf{y}_{1:(n-1)},C_{n-1}=t_i \mid \hat \gamma,\hat \eta).
     \end{aligned}
     \label{equ: joint_dist_algorithm_1}
        \end{equation}

    
    When $t_n$ is a changepoint, we have $C_n = t_n$. Then 
	    \begin{equation}
     \begin{aligned}
         &p(\mathbf{y}_{1:n}, C_n=t_n \mid \hat \gamma,\hat \eta) = p(y_{n} \mid \hat \gamma,\hat \eta) H(t_n)\sum_{j=\hat{j}}^{n-1}  p(\mathbf{y}_{1:(n-1)},C_{n-1}=t_j \mid \hat \gamma,\hat \eta).
     \end{aligned}
     \label{equ: joint_dist_algorithm_2}
        \end{equation}
    \item \textbf{Determine the most recent changepoint by}
    \begin{equation}
    \begin{aligned}
       \hat{C}_n & = \argmax_{t_{\hat{j}}\leq t_i\leq t_n} p(\mathbf{y}_{1:n}, C_n=t_i \mid \hat \gamma,\hat \eta).
    \end{aligned}
\end{equation}
	
	
	\end{enumerate} 
	\label{algm: algm 1}
\end{algorithm}



We summarize our approach in Algorithm \ref{algm: algm 1} for detecting the most recent changepoint.  First, we apply Theorem \ref{thm: KF update} multiple times to obtain the sequence of predictive distributions, i.e., $p(y_{n} \mid \mathbf{y}_{i:(n-1)}, \gamma, \eta)$ for $i=1, \dots, n-1$. Next, given the predictive distributions, we compute the joint distribution $p(\mathbf y_{1:n}, C_n=t_i)$ using Equations (\ref{equ: joint_dist_algorithm_1}) and (\ref{equ: joint_dist_algorithm_2}). Finally, we estimate $\hat{C}_n$, the most recent changepoint before or at time $t_n$, by 
the MAP of the joint distribution, i.e., $\hat{C}_n = \argmax_{t_1\leq t_i\leq t_n} p(\mathbf{y}_{1:n}, C_n=t_i)$.  

{We call the online changepoint detection approach in Algorithm \ref{algm: algm 1}   the \textit{sequential Kalman filter} (SKF) because in Step 2, we sequentially compute the predictive distribution $p(y_{n} \mid \mathbf{y}_{i:(n-1)}, \gamma, \eta)$ for $i=1, \dots, n-1$ using the KF, and employ them  for computing the joint distributions in Equations (\ref{equ: joint_dist_algorithm_1}) and (\ref{equ: joint_dist_algorithm_2}) in Step 3. This sequential approach iterates over different starting values of time index $i$ by stitching different KFs together.  
}

\subsection{Computational Complexity}
\label{subsec: computation_complexity}

\begin{figure}[]
\centering
\includegraphics[scale=0.7]{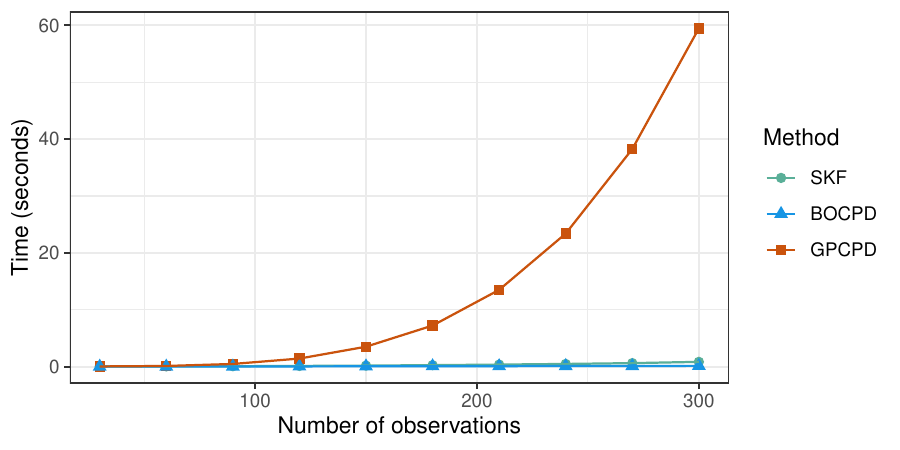}
\caption{The comparison of the computational cost between the SKF, BOCPD, and GPCPD methods.}
\label{fig: time complex}
\end{figure}

Let $n$ denote the total number of observations. When there is no changepoint detected before, the SKF algorithm requires $\mathcal{O}(n)$ operations at time $t_n$ by computing $n-1$ predictive distributions $p(y_n \mid \mathbf{y}_{i:(n-1)})$ for $i=1,\dots,n-1$, each taking $O(1)$ operation according to Theorem \ref{thm: KF update}. When there is at least one changepoint detected before, by applying the truncation approach described in Section \ref{sec: background},  we only need to compute predictive distributions $p(y_n \mid \mathbf{y}_{i:(n-1)})$ for $i=\hat{C}_{n-1},\dots,n-1$, which reduce the computational complexity to $\mathcal{O}(n^{\prime})$, where $n^{\prime} = n - \hat{C}_{n-1}$, and $\hat{C}_{n-1}$ denotes the most recently detected changepoint before the time $t_{n-1}$. When there is no changepoint detected before, we have $n'=n$. 

In Figure \ref{fig: time complex}, we compare the computational time for three distinct methods as the number of observations increases on a Windows 10 PC with two 3.00GHz i7-9700 CPUs. The computational cost of SKF  is substantially smaller than  GPCPD, as direct computation requires inversion of the covariance matrix. The fast computation enables us to deploy the scalable SKF algorithm for real-world scenarios with a large number of observations. 
On the other hand, the cost of  SKF  is similar to that of BOCPD \citep{fearnhead2007line,adams2007bayesian}, whereas the temporal correlation is modeled in  SKF  but not in BOCPD. As temporal correlation widely exists in real-world data sets, modeling the correlation can improve the accuracy of the changepoint detection. More detailed comparisons with other methods are provided in Section \ref{sec: comparison} in the supplementary material. 

\section{Simulation Studies}
\label{sec: simulation}

This section compares different approaches for estimating single and multiple changepoints from temporally correlated data.  We consider three types of changes: mean, variance, and correlation. {For initial states before the changes, the data is sampled from a Gaussian process with mean $\mu = 0$, variance $\sigma^2=1$, and nugget parameter  $\eta=\frac{\sigma_0^2}{\sigma^2}=0.1$. We employ covariance functions from Equations (\ref{equ: exp_kernel}) and (\ref{equ: Matern_kernel}) in simulations, setting range parameters at $\gamma=12$ and $\gamma=4$, respectively. The parameters $\mu$, $\sigma$, and $\gamma$ can vary under different change scenarios specified later.}
We compare the SKF approach with the BOCPD approach  \citep{fearnhead2007line,adams2007bayesian} and CUSUM algorithm \citep{page1954continuous} summarized in Section \ref{sec: cusum} of the supplementary material. 
For the SKF algorithm, the range and nugget parameters in the covariance matrix are estimated by maximizing the marginal likelihood function  in (\ref{equ:mle}) 
for computing the predictive distributions. 
In Section \ref{sec: misspecified_settings}  of the supplementary material, we also compare different approaches for the scenarios when a covariance function is misspecified, and the conclusion is in line with the results herein. 


\subsection{Single Changepoint}
\label{sec: single cp}

We first compare the performance of different approaches for time sequences with a single changepoint. 
In this scenario, each method can report at most one changepoint during the whole detection period and thus only the first detected changepoint will be recorded. 
We apply two commonly used metrics for online changepoint detection algorithms \citep{basseville2014sequential, chen2019sequential}  to evaluate the performance of each method: Average Detection Delay (ADD) and Average Run Length (ARL). 
The ADD, defined as $\mathbb{E}_{\tau}\left[(\Gamma - \tau)^+\right]$, where the metric $\Gamma$ represents the earliest time we detect a changepoint around the latent changepoint $\tau$. 
ADD measures the average time lag between a changepoint occurrence and the time of its first detection, which may be compared with the power of a statistical test in hypothesis testing. A small value of the ADD indicates that the method is more powerful in detecting a latent changepoint.  The ARL, defined as $\mathbb{E}_{\infty} \left[\Gamma\right]$, measures the average time of the first detected changepoint when there is no changepoint in the data, which can be interpreted as the type-I error in hypothesis testing. 
We evaluate SKF, BOCPD, and CUSUM using 100 random sampled time series, each with $n=100$ observations.
The observations are equally spaced in time, where the first $n_0=50$ observations serve as the training samples. 
To ensure a fair comparison, we let the ARL be approximately 50 across all methods, through specifying the hazard parameter value in BOCPD or SKF, and the threshold value for the CUSUM method, based on the time period with no changepoint.

\begin{figure}[t]
\centering
\includegraphics[scale=0.7]{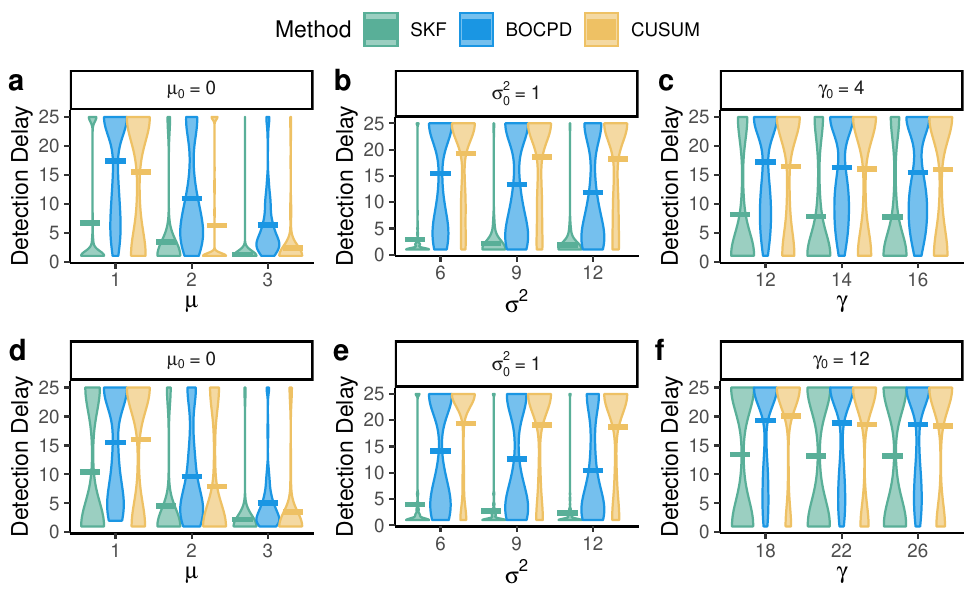}
\caption{Violin plots comparing average detection delay for SKF, BOCPD, and CUSUM methods for 100 simulations. The upper and lower panels show the detection delay of each method when the data are simulated with the Mat{\'e}rn correlation with the roughness parameter being 2.5 and exponential correlation, respectively. A method with a low average detection delay is better.
$\mu_0$, $\sigma_0^2$, and $\gamma_0$ represent pre-change parameter values, while $\mu$, $\sigma^2$ and $\gamma$ on the x-axis stand for post-change parameter values.
}  
\label{fig: sim_single_cp}
\end{figure}

Figure \ref{fig: sim_single_cp} shows that SKF consistently outperforms BOCPD and CUSUM in all tested scenarios, achieving the lowest ADD. 
In particular, when the change contains a small mean shift, a variance or correlation shift, both CUSUM and BOCPD have a large delay in detecting, whereas the SKF method has a relatively low detection delay for all scenarios. 
 The SKF performs better than other approaches as it captures  the temporal correlation from the observations. Furthermore, the computational complexity of the SKF  is similar to BOCPD,  which is crucial for real-world applications with a large number of samples. 
 
 {Additionally, in Section \ref{sec: misspecified_settings} of the supplementary material, we examine the SKF  with a misspecified covariance function. 
 Figure \ref{fig: mis_add_comparison} demonstrates that even with the misspecified covariance, the SKF  performs comparably well to scenarios with the correct covariance, and still outperforms the BOCPD. 
 This is because a method having a misspecified covariance with an estimated correlation length scale parameter is typically better than assuming the independence between observations to approximate the temporal covariance in the underlying data-generating process. 
 This result reveals the robustness of the SKF  for changepoint detection, even when certain configurations deviate from the true data-generating process. 
 }

\subsection{Multiple Changepoints}
\label{sec: multiple changepoints}

In this section, we assess the performance of different algorithms to detect multiple changepoints. We sample 100 time sequences, each containing $n=150$ observations from a GP having the Mat{\'e}rn covariance in (\ref{equ: Matern_kernel}) for demonstration purposes. Each time series have four changepoints at positions $\bm \tau = \{33, 66, 98, 130\}$. We investigate three scenarios where the changes occur in mean, variance, and covariance range parameters. 

In multiple changepoints scenarios, metrics like ADD and ARL are not suitable as the number of detected and true changepoints  may not be the same. Thus, we use the covering metric \citep{arbelaez2010contour, van2020evaluation} that measures how well the detected changepoints align with the true changepoints, defined in Section \ref{sec: define_covering} of the supplementary material. A method with a larger value of the covering metric is better.  

\begin{figure}[t]
\centering
\includegraphics[scale=0.7]{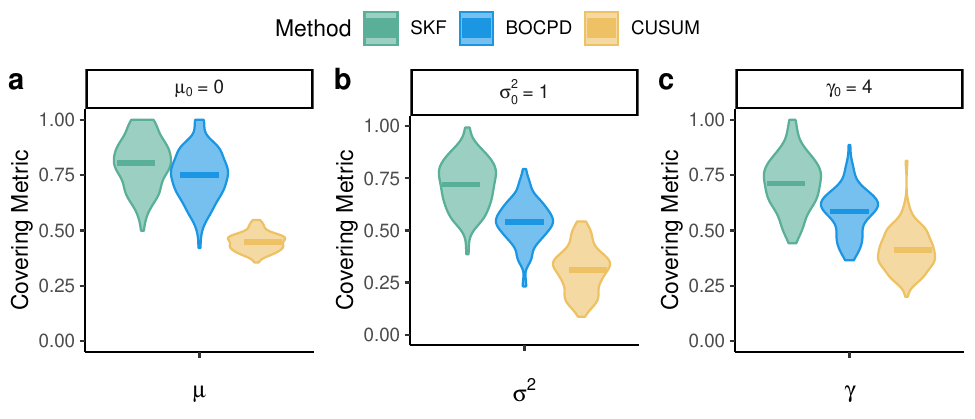}
\caption{Violin plots of the covering metric (larger values are better) of the SKF, BOCPD, and CUSUM methods for simulated data with multiple changepoints. 
}
\label{fig: cover_multiple_cp}
\end{figure}

 Panels a-c in Figure \ref{fig: cover_multiple_cp} show the average covering metric for SKF, BOCPD, and CUSUM methods for the scenarios with the mean, variance, and correlation changes, respectively.  
{Both  BOCPD and SKF approaches outperform the CUSUM method in terms of the covering metric across all scenarios. This is because the CUSUM method relies on a prespecified threshold of the test statistics written as a cumulative summation of information to determine whether the current time point is a  changepoint, which does not look back to find a changepoint in prior time points as BOCPD and SKF. 
In contrast, the predictive distributions in BOCPD and SKF contain information from a time period of previous subsequences, which enables the methods to detect a changepoint when information accumulates.  
} 
 {Furthermore, the SKF method outperforms both BOCPD and CUSUM methods in terms of the covering metric, as the temporal correlations from the data are modeled in SKF, making  SKF more accurate to approximate the data-generating mechanism.}

\begin{figure}[t]
    \centering 
    \includegraphics[scale=.6,trim=10 0 0 10,clip]{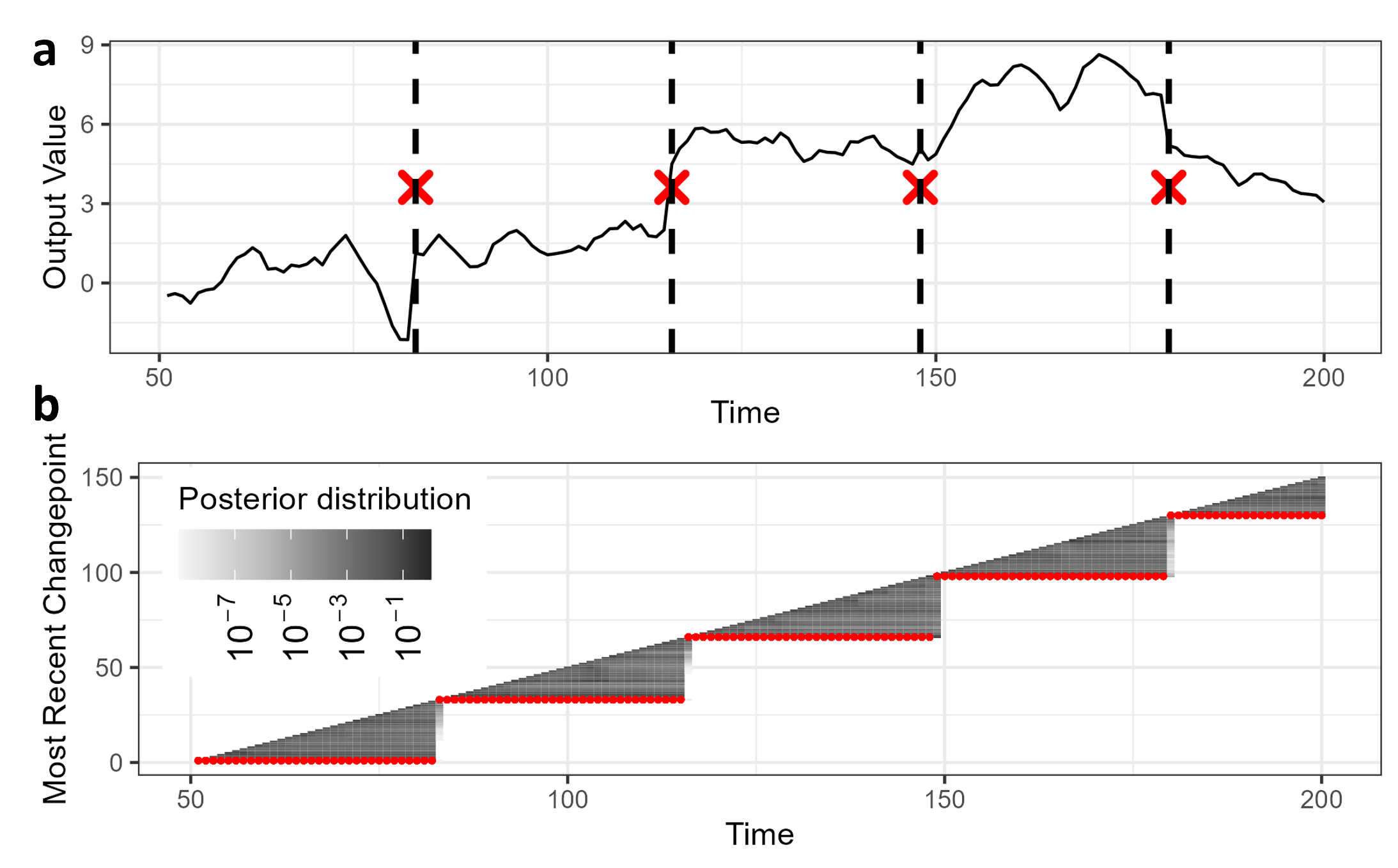}
    \caption{The black curves in panel a give the temporally correlated outcomes with 4 mean changes. The black dashed lines indicate the true changepoint locations and the red crosses give the estimated changepoints by SKF. 
    Panel b shows the posterior distribution of the most recent changepoints at each time point, with  MAP  estimates graphed as  red dots. }
    \label{fig:multiple cps}
\end{figure}

 Panel a in Figure \ref{fig:multiple cps}  graphs the detected changepoints by SKF and the true changepoint for a simulated case with the mean shift.  Panel b  gives the classification probability of the most recent changepoint at each time point.  The estimated most recent changepoints, marked by the red solid points, are the ones with the maximum posterior probability over all possible values.
 A new changepoint is identified if the most recently detected changepoint differs from the previously detected changepoint.

\section{SARS-CoV-2 Detection among Dialysis Patients}
\label{sec: COVID detection}

\begin{figure}[t]
    \centering
    \includegraphics[scale=0.55] 
    {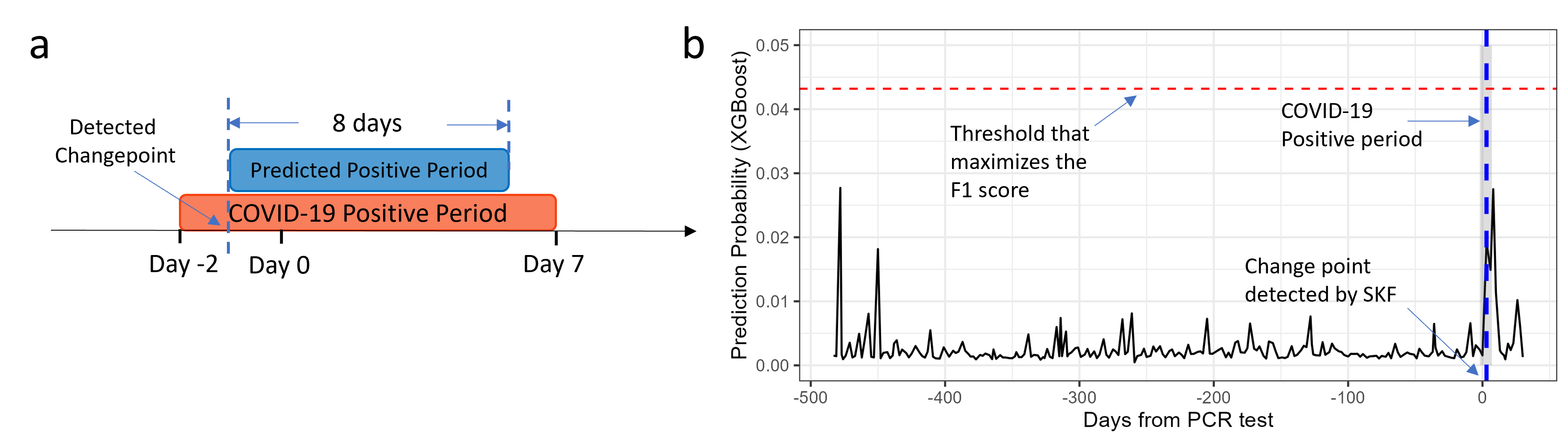}
    \caption{In panel a, the orange area is the COVID-19 positive period spanning from day -2 to day 7, where a patient has a positive COVID-19 PCR test at day 0. 
    If a patient is detected to be COVID-19 positive, the detected changepoint and the subsequent seven days are marked as the predicted infection period, shown as the blue area. The black curve in panel b shows the probability sequence of being COVID-19 positive estimated by XGBoost, and a value larger than the threshold value shown as
    red dashed line is classified as COVID-19 positive. The blue dashed line marks the changepoint detected by SKF, and the grey area represents the COVID-19 positive period from day -2 to day 7. }
    \label{fig: COVID_CPD_Patient}
\end{figure}

We focus on detecting COVID-19 infection for a large number of dialysis patients in this section. Data-driven models have been extensively used for detecting 
COVID-19 infection dates in patient-level longitudinal data \citep{li2020using, zoabi2021machine, monaghan2021machine}. Most of these models generate prediction probabilities of COVID-19 infections, whereas a threshold is typically used to identify infections, and the temporal pattern among the longitudinal data was not modeled in these approaches. {For example, some COVID-19 positive patients have only mild to moderate symptoms, which may result in an increasing trend of the prediction probabilities, but the prediction probabilities might still fall below the pre-specified threshold. Ignoring such a trend can lead to low sensitivity in detection.}
We will apply changepoint detection methods to probability sequences of COVID-19  infections to improve the detection performance. 

This study analyzes longitudinal treatment data from over 150,000 dialysis patients collected by Fresenius Kidney Care  between January 2020 and March 2022. Each patient visits the clinics about three times per week, producing a large data set with millions of observations. For each clinic visit, the data includes features such as sitting blood pressure, weight, temperature, respiration rate, pulse rate, oxygen level, interdialytic weight gain, average blood flow rate, and average dialysis flow rate. 
The dataset contains 15 million samples, 
with each patient owning around 94 samples on average, 
where only $0.4\%$ of the observations are labeled as COVID-19 positive.  {We give an example of the mechanism of detection in  Panel a Figure \ref{fig: COVID_CPD_Patient}, where the COVID-19 positive window of a patient contains a three-day incubation period  \citep{jansen_2021_investigation, song_2022_serial} and a seven-day infection period post symptom onset \citep{Hakki2022Onset}. 
 A clinic visit is labeled as COVID-19 positive if it is within two days prior to, or seven days following day 0, which is the day for a positive COVID-19 PCR test.  We conducted the sensitivity analysis with different choices of COVID-19 positive period in Section
\ref{sec: sensitivity analysis} in the supplementary material and the results remain similar.}

{We employed an integrated procedure to detect the changepoint from the COVID-19 infection summarized in Algorithm \ref{algm: algm 2} of the supplementary material. We first apply a data-driven classification model to patients' clinical data, here chosen as the XGBoost method \citep{chen2016xgboost}, which was previously found to be accurate in detecting COVID-19 among dialysis patients \citep{monaghan2021machine,Juntao2023Predicting} compared to a few other classification methods. Second, we apply SKF 
to detect the change in the daily prediction probabilities of COVID-19 infection from the XGBoost approach.  Furthermore,  we developed an additional screening step to detect the onset of an increasing subsequence in infection probabilities through a hypothesis test (Step 5 in Algorithm \ref{algm: algm 2}), as typically the increase of the probability sequences of infection should be detected. 
Once the detected changepoint passes this screening step, we mark the seven-day period after the detected changepoint as COVID-19 positive
\citep{Hakki2022Onset}.  
The integrated approach is generally applicable to detect changes from longitudinal data. Details of the integrated approach can be found in Section \ref{sec: derivation_test_statistic} of the supplementary material. In both BOCPD and SKF methods, the hazard function was defined in proportion to the county-level daily probability of contracting COVID-19 \citep{li2021robust}, enhancing detection accuracy by the estimated daily transmission probability based on local infection and death counts.}

{
A typical COVID-19 positive patient will have two changes during the infection period, characterized by an increasing trend at the beginning and a decreasing trend at the end of the infection probability.  We aim to detect the first change, and such a detection scheme is useful for the onset of other diseases based on longitudinal data. This means the covering metric in Section \ref{sec: multiple changepoints} is not sensible.   To better evaluate the effectiveness of detecting the infection, we use precision, recall, F1-score, and detection delay as our performance metrics, which are defined below. 
}
The threshold value for the classification probabilities from the XGBoost method is determined by maximizing the F1-score across all patients in the test data, which is defined as the harmonic mean of precision and recall:  $  \text{F1}\text{-score} = 2 \times \frac{\text{precision} \times \text{recall}}{(\text{precision} +  \text{recall})}$, 
where $\text{precision} = \frac{\text{TP}}{\text{TP} +  \text{FP}}$ is the ratio of the true positives out of all the positive predictions, and $\text{recall} = \frac{\text{TP}}{\text{TP} +  \text{FN}}$ is ratios of true positive out of all the positively labeled samples, which quantifies the power of the algorithm.
The true positives (TP), false positives (FP), and false negatives (FN) are defined as $      \mbox{TP} = \sum_{j=1}^{n^*} I_{\{\hat{x}_j=1, x_j=1\}},~ \mbox{FP} = \sum_{j=1}^{n^*} I_{\{\hat{x}_j=1, x_i=0\}}$, and $\mbox{FN} = \sum_{j=1}^{n^*} I_{\{\hat{x}_j=0, x_j=1\}} $, 
where $n^* = n-n_0$ denotes the number of samples in the test data. $I_{\{\cdot\}}$ is the indicator function, $x_j$ are the actual COVID-19 labels and $\hat{x}_j$ is the predictive COVID-19 labels. 
{For statistical learning models like XGBoost, the predicted label $\hat{x}_j$ is assigned a value of $1$ if $t_j$ is within a seven-day window following the date when the predicted probability exceeds the threshold value, and is set to $0$ otherwise.} For changepoint detection algorithms such as SKF and BOCPD, $\hat{x}_j$ is assigned the value of $1$ if $t_j$ falls within a seven-day period following a detected changepoint. 
 The average detection delay is calculated as the average number of days from the start date of the COVID-19 positive period to the time a changepoint within the positive period is first detected.  A lower average detection delay reflects a quicker response to the onset of changes. Furthermore,  any detection made after 2 weeks of day 0 is considered as not useful in the online detection, which is not counted as a true positive, as they are too late to help. This threshold can be adjusted for detecting other diseases. 


\begin{table}
\centering
\caption{Out-of-sample comparisons of the classification methods and online changepoint detection methods, including CUSUM, BOCPD, and SKF on COVID-19 detection with the baseline positive rate of $0.4\%$.}
\label{tbl: compare_SKF_threshold}
\begin{tabular}{|c|c|c|c|c|} 
\hline
                                      & Precision                  & Recall                     & F1-score                   & Detection Delay            \\ 
\hline
Logistic regression                   & 0.055                      & 0.141                      & 0.079                      & 1.56                          \\ 
\hline
Random forests                         & 0.083                      & 0.123                      & 0.099                      & 2.15                          \\ 
\hline
XGBoost                               & 0.082                      & 0.179                      & 0.113                      & 2.22                         \\ 
\hline
CUSUM                                 & 0.028                      & 0.023                      & 0.025                      & 3.73                       \\ 
\hline
BOCPD                                 & 0.174                      & 0.168                      & 0.171                      & 5.3                        \\ 
\hline
SKF                                   & 0.218                      & 0.179                      & 0.197                      & 4.13                       \\ 
\hline
SKF with screening & 0.232 &0.190 & 0.209 & 2.82  \\
\hline
\end{tabular}
\end{table}
Table \ref{tbl: compare_SKF_threshold} compares the performance of different approaches for detecting COVID-19 infection. 
Here the baseline positive data only constitutes around $0.4\%$ of the total observations. This setting differs from a few other COVID-19 detection schemes where each COVID-19 positive record is matched with a few negative samples \citep{monaghan2021machine, Juntao2023Predicting}. The practical scheme is closer to the longitudinal detection scheme employed herein. 
We first found that changepoint detection algorithms, including BOCPD and SKF, perform better than the classification methods, such as logistic regression, random forecast, and XGBoost. This is because the change detection utilizes longitudinal information for identifying the change for each patient, while the classification methods rely on a unified threshold of probability sequences of being infected for all patients. Among the changepoint detection methods, the CUSUM algorithm is not as good as BOCPD and SKF. The good performance of SKF and BOCPD is largely due to their ability to recursively inspect whether each of the previous time points is a changepoint, in contrast to the CUSUM method which can only determine whether the current time point is a  changepoint.
{Second, SKF outperforms BOCPD in both F1-score and detection delay. This advantage is largely attributed to SKF's ability to model temporal correlations, which helps reduce false detection. Notably, SKF can detect the infection about one day faster than BOCPD on average, which means the SKF requires less information to identify a COVID-19 infection. Furthermore, incorporating the screening method, as detailed in Steps 6 and 7 of Algorithm \ref{algm: algm 2} in the supplementary material, further enhances the F-1 score and reduces detection delay in SKF. This improvement aligns with our expectations, since the screening method chooses changepoints related to an increasing subsequence, thereby improving the precision of COVID-19 detection.} 

\begin{figure}[t]
    \centering
    \includegraphics[scale=.7]{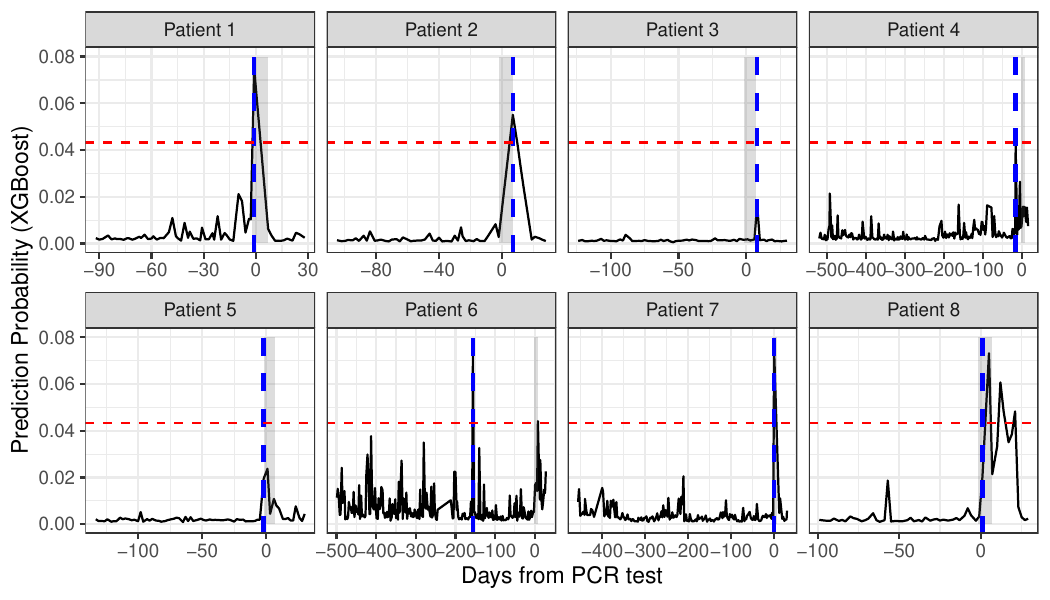}
    \caption{Results of SKF on the probability sequences of 16 COVID-19 patients. The red dashed line is the threshold value that maximizes the F1-score. The blue dashed line shows the changepoint detected by SKF. The grey area is the period that a patient is labeled as COVID-19-positive. Day 0 is the positive PCR test day.}
    \label{fig: COVID_CPD_Patient_group}
\end{figure}

Panel b in Figure \ref{fig: COVID_CPD_Patient} gives an example comparing the SKF detection with the XGBoost method. The probability sequences of COVID-19 infection from XGBoost, shown as the black curve, consistently remain beneath the threshold value, indicated by the red dashed line, suggesting that this patient is predicted as COVID-19 negative by XGBoost for the entire period. 
The increasing trend of probability sequence during the infection period, however, allows the SKF  to successfully detect the changepoint from COVID-19 infection, indicated by the blue dashed line. {Figure \ref{fig: COVID_CPD_Patient_group} gives further comparison between the SKF and XGBoost for a group of 8 randomly selected COVID-19 positive patients. Here we only show the plots of positive patients and there are around $74\%$ of the patients who do not have a positive PCR test during the whole period. The SKF method successfully identifies the COVID-19 positive period for 6 of these patients and misses the COVID-19 infection for 2 patients. In comparison, the XGBoost method correctly identifies the COVID-19 positive period for only 4 patients and misses the infection for 4 patients.} 
SKF may be preferred for this problem over a conventional classification approach as it can identify the probability subsequences with an increasing trend. 
Furthermore,  the empirical autocorrelation of probability sequences of a few randomly selected patients is plotted in Section \ref{supsec: cor in COVID paitent} in the supplementary material. The autocorrelation of the probability sequences from the longitudinal data is modeled in SKF, which improves the detection accuracy. More numerical comparisons of online changepoint detection approaches for a few other real-world examples are provided in Section \ref{supsec: real data analysis} in the supplementary material, which confirms competitive performance by the SKF approach.

\section{Conclusion}
\label{sec: conclusion}

This paper introduces the Sequential Kalman Filter (SKF) for online changepoint detection for data with temporal correlations. 
 The temporal correlation between each time point is modeled in SKF and  
the computational cost is dramatically reduced without approximating the likelihood function. 
Furthermore, we developed a new approach that integrates high-dimensional covariates and massive outcomes for detecting COVID-19 infection from a large longitudinal dataset of dialysis patients, overcoming the challenge of modeling massive longitudinal covariates with a large proportion of missingness. 
The new approach 
substantially improves detection accuracy compared to conventional classification and other online changepoint detection approaches. 




We outline a few research directions following this work. 
Online changepoint detection of a large number of
time series 
is common for longitudinal health records and spatio-temporal climate data \citep{chang2016changes}.   
While principal component analysis was used  before the CUSUM algorithm for changepoint detection of multivariate time series  \citep{kurt2020real}, 
 coherent statistical models that consider temporal correlation, such as vector autoregressive models \citep{prado2010time} and latent factor processes \citep{gu2022gaussian}, may be appealing to extend the scalable changepoint detection approach. Distributed inference that maintains efficiency while not requiring whole patient information from different clinics or hospital is of interest \citep{duan2022heterogeneity}, whereas temporal correlation from observations was not often modeled in this scenario. 
Furthermore, shrinkage estimators were studied for inducing sparsity structure of the state space model for estimating Granger causality \citep{shojaie2022granger}, whereas efficient models of the noise and scalable estimation of the change of the dynamics were not studied.

\bigskip
\begin{center}
{\large\bf SUPPLEMENTARY MATERIAL}
\end{center}

\begin{description}

\item[R package for the SKF algorithm:] R-package \texttt{SKFCPD} that efficiently implements the SKF algorithm on both the univariate and multivariate time series with acceptance of missing values.

\item[Supplementary Material:] In Section \ref{supsec: cor in COVID paitent}, we present the autocorrelation and partial autocorrelation 
for probability sequences of four dialysis patients. The derivation of Equation (\ref{equ: joint_dist_BOCPD_3}) is provided in Section \ref{supsec: proof_of_equation_3}. Section \ref{sec: connection_DLM_and_Matern} investigates the relationships between the DLM and the GP model with the Mat{\' e}rn covariance in Equation (\ref{equ: Matern_kernel}).
The derivation of Equation (\ref{equ: pred dist model 2}) can be found in Section \ref{sec: Proof equ: pred dist model 2}. Section \ref{sec: KF_for_DLM_app} summarizes the KF for DLM. The proofs of Lemma \ref{lemma:a_b_k} and Theorem \ref{thm: KF update} are included in Section \ref{sec: proof_of_Lemma_1} and \ref{sec: proof_thm_1}, respectively. Section \ref{sec: comparison} compares the SKF algorithm with other models. Section \ref{sec: cusum} reviews the CUSUM method. Section \ref{sec: misspecified_settings} examines the SKF algorithm with misspecified configurations. Section \ref{sec: define_covering} provides the definition of the covering metric. Section \ref{sec: derivation_test_statistic} shows the derivation of the test statistic in Algorithm \ref{algm: algm 2}. Section \ref{sec: sensitivity analysis} gives the sensitivity analysis for the COVID-19 positive period. Finally, Section \ref{supsec: real data analysis} features a comparative analysis between the SKF algorithm and the BOCPD method using real benchmark datasets. 
\end{description}

\bigskip

\if0\blind
{
\begin{center}
{\large\bf ACKNOWLEDGEMENTS}
\end{center}

This research is partially supported
by NSF under award number 2053423 and NIH under award number R01DK130067. We thank Fresenius Medical Care North America for providing de-identified data of health records for dialysis patients.
} \fi

\if0\blind
{
\begin{center}
{\large\bf DISCLOSURE STATEMENT}
\end{center}

The authors report there are no competing interests to declare.
} \fi

\bibliographystyle{chicago}

\bibliography{Bibliography.bib}

\begin{thebibliography}{}

\bibitem[\protect\citeauthoryear{Adams and MacKay}{Adams and
  MacKay}{2007}]{adams2007bayesian}
Adams, R.~P. and D.~J. MacKay (2007).
\newblock Bayesian online changepoint detection.
\newblock {\em arXiv preprint arXiv:0710.3742\/}.

\bibitem[\protect\citeauthoryear{Arbelaez, Maire, Fowlkes, and Malik}{Arbelaez
  et~al.}{2010}]{arbelaez2010contour}
Arbelaez, P., M.~Maire, C.~Fowlkes, and J.~Malik (2010).
\newblock Contour detection and hierarchical image segmentation.
\newblock {\em IEEE transactions on pattern analysis and machine
  intelligence\/}~{\em 33\/}(5), 898--916.

\bibitem[\protect\citeauthoryear{Basseville, Nikiforov, and
  Tartakovsky}{Basseville et~al.}{2014}]{basseville2014sequential}
Basseville, M., I.~Nikiforov, and A.~Tartakovsky (2014).
\newblock {\em Sequential Analysis: Hypothesis Testing and Changepoint
  Detection}.
\newblock CRC Press.

\bibitem[\protect\citeauthoryear{Basseville and Nikiforov}{Basseville and
  Nikiforov}{1993}]{basseville1993detection}
Basseville, M. and I.~V. Nikiforov (1993).
\newblock {\em Detection of abrupt changes : theory and applications}, Volume
  104.
\newblock Englewood Cliffs (New Jersey): Prentice-Hall.

\bibitem[\protect\citeauthoryear{Chang, Stein, Wang, Kotamarthi, and
  Moyer}{Chang et~al.}{2016}]{chang2016changes}
Chang, W., M.~L. Stein, J.~Wang, V.~R. Kotamarthi, and E.~J. Moyer (2016).
\newblock Changes in spatiotemporal precipitation patterns in changing climate
  conditions.
\newblock {\em Journal of Climate\/}~{\em 29\/}(23), 8355--8376.

\bibitem[\protect\citeauthoryear{Chen}{Chen}{2019}]{chen2019sequential}
Chen, H. (2019).
\newblock Sequential change-point detection based on nearest neighbors.
\newblock {\em The Annals of Statistics\/}~{\em 47\/}(3), 1381--1407.

\bibitem[\protect\citeauthoryear{Chen and Guestrin}{Chen and
  Guestrin}{2016}]{chen2016xgboost}
Chen, T. and C.~Guestrin (2016).
\newblock {XGB}oost: A scalable tree boosting system.
\newblock In {\em Proceedings of the 22nd acm sigkdd international conference
  on knowledge discovery and data mining}, pp.\  785--794.

\bibitem[\protect\citeauthoryear{Chen, Banerjee, Dom{\'i}nguez-García, and
  Veeravalli}{Chen et~al.}{2016}]{chen2016quickest}
Chen, Y.~C., T.~Banerjee, A.~D. Dom{\'i}nguez-García, and V.~V. Veeravalli
  (2016).
\newblock Quickest line outage detection and identification.
\newblock {\em IEEE Transactions on Power Systems\/}~{\em 31\/}(1), 749--758.

\bibitem[\protect\citeauthoryear{Chu and Chen}{Chu and
  Chen}{2019}]{hao2019asympotic}
Chu, L. and H.~Chen (2019).
\newblock {Asymptotic distribution-free change-point detection for multivariate
  and non-Euclidean data}.
\newblock {\em The Annals of Statistics\/}~{\em 47\/}(1), 382 -- 414.

\bibitem[\protect\citeauthoryear{Doucet, Godsill, and Andrieu}{Doucet
  et~al.}{2000}]{doucet2000sequential}
Doucet, A., S.~Godsill, and C.~Andrieu (2000).
\newblock On sequential {M}onte {C}arlo sampling methods for {B}ayesian
  filtering.
\newblock {\em Statistics and computing\/}~{\em 10}, 197--208.

\bibitem[\protect\citeauthoryear{Duan, Li, Ma, Zhang, Lasky, Monaghan,
  Chaudhuri, Usvyat, Gu, Guo, Kotanko, and Wang}{Duan
  et~al.}{2023}]{Juntao2023Predicting}
Duan, J., H.~Li, X.~Ma, H.~Zhang, R.~Lasky, C.~K. Monaghan, S.~Chaudhuri, L.~A.
  Usvyat, M.~Gu, W.~Guo, P.~Kotanko, and Y.~Wang (2023).
\newblock Predicting {SARS-CoV}-2 infection among hemodialysis patients using
  multimodal data.
\newblock {\em Frontiers in Nephrology\/}~{\em 3}, 1179342.

\bibitem[\protect\citeauthoryear{Duan, Ning, and Chen}{Duan
  et~al.}{2022}]{duan2022heterogeneity}
Duan, R., Y.~Ning, and Y.~Chen (2022).
\newblock Heterogeneity-aware and communication-efficient distributed
  statistical inference.
\newblock {\em Biometrika\/}~{\em 109\/}(1), 67--83.

\bibitem[\protect\citeauthoryear{Durbin and Koopman}{Durbin and
  Koopman}{2012}]{durbin2012time}
Durbin, J. and S.~J. Koopman (2012).
\newblock {\em Time series analysis by state space methods}, Volume~38.
\newblock OUP Oxford.

\bibitem[\protect\citeauthoryear{Fearnhead}{Fearnhead}{2006}]{fearnhead2006exact}
Fearnhead, P. (2006).
\newblock Exact and efficient {B}ayesian inference for multiple changepoint
  problems.
\newblock {\em Statistics and computing\/}~{\em 16\/}(2), 203--213.

\bibitem[\protect\citeauthoryear{Fearnhead and Liu}{Fearnhead and
  Liu}{2007}]{fearnhead2007line}
Fearnhead, P. and Z.~Liu (2007).
\newblock On-line inference for multiple changepoint problems.
\newblock {\em Journal of the Royal Statistical Society: Series B (Statistical
  Methodology)\/}~{\em 69\/}(4), 589--605.

\bibitem[\protect\citeauthoryear{Fearnhead and Liu}{Fearnhead and
  Liu}{2011}]{fearnhead2011efficient}
Fearnhead, P. and Z.~Liu (2011).
\newblock Efficient {B}ayesian analysis of multiple changepoint models with
  dependence across segments.
\newblock {\em Statistics and Computing\/}~{\em 21\/}(2), 217--229.

\bibitem[\protect\citeauthoryear{Fryzlewicz}{Fryzlewicz}{2014}]{fryz2014wild}
Fryzlewicz, P. (2014).
\newblock {Wild binary segmentation for multiple change-point detection}.
\newblock {\em The Annals of Statistics\/}~{\em 42\/}(6), 2243 -- 2281.

\bibitem[\protect\citeauthoryear{Gu and Li}{Gu and Li}{2022}]{gu2022gaussian}
Gu, M. and H.~Li (2022).
\newblock Gaussian orthogonal latent factor processes for large incomplete
  matrices of correlated data.
\newblock {\em Bayesian Analysis\/}~{\em 17\/}(4), 1219--1244.

\bibitem[\protect\citeauthoryear{Gu, Liu, Fang, and Tang}{Gu
  et~al.}{2023}]{gu2022scalable}
Gu, M., X.~Liu, X.~Fang, and S.~Tang (2023).
\newblock {Scalable Marginalization of Correlated Latent Variables with
  Applications to Learning Particle Interaction Kernels}.
\newblock {\em The New England Journal of Statistics in Data Science\/}, 1--15.

\bibitem[\protect\citeauthoryear{Hakki, Zhou, Jonnerby, Singanayagam, Barnett,
  Madon, Koycheva, Kelly, Houston, Nevin, Fenn, Kundu, Crone, Pillay, Ahmad,
  Derqui-Fernandez, Conibear, Freemont, Taylor, and Ferguson}{Hakki
  et~al.}{2022}]{Hakki2022Onset}
Hakki, S., J.~Zhou, J.~Jonnerby, A.~Singanayagam, J.~L. Barnett, K.~J. Madon,
  A.~Koycheva, C.~Kelly, H.~Houston, S.~Nevin, J.~Fenn, R.~Kundu, M.~A. Crone,
  T.~D. Pillay, S.~Ahmad, N.~Derqui-Fernandez, E.~Conibear, P.~S. Freemont,
  G.~P. Taylor, and N.~Ferguson (2022).
\newblock Onset and window of {SARS-CoV}-2 infectiousness and temporal
  correlation with symptom onset: a prospective, longitudinal, community cohort
  study.
\newblock {\em The Lancet Respiratory Medicine\/}~{\em 10\/}(11), 1061–1073.

\bibitem[\protect\citeauthoryear{Handcock and Stein}{Handcock and
  Stein}{1993}]{handcock1993bayesian}
Handcock, M.~S. and M.~L. Stein (1993).
\newblock A {B}ayesian analysis of kriging.
\newblock {\em Technometrics\/}~{\em 35\/}(4), 403--410.

\bibitem[\protect\citeauthoryear{Harris, Li, and Tucker}{Harris
  et~al.}{2022}]{harris2022scalable}
Harris, T., B.~Li, and J.~D. Tucker (2022).
\newblock Scalable multiple changepoint detection for functional data
  sequences.
\newblock {\em Environmetrics\/}~{\em 33\/}(2), e2710.

\bibitem[\protect\citeauthoryear{Hartikainen and S{\"a}rkk{\"a}}{Hartikainen
  and S{\"a}rkk{\"a}}{2010}]{hartikainen2010kalman}
Hartikainen, J. and S.~S{\"a}rkk{\"a} (2010).
\newblock Kalman filtering and smoothing solutions to temporal {G}aussian
  process regression models.
\newblock In {\em 2010 IEEE international workshop on machine learning for
  signal processing}, pp.\  379--384. IEEE.

\bibitem[\protect\citeauthoryear{Haynes, Fearnhead, and Eckley}{Haynes
  et~al.}{2017}]{haynes2017computationally}
Haynes, K., P.~Fearnhead, and I.~A. Eckley (2017).
\newblock A computationally efficient nonparametric approach for changepoint
  detection.
\newblock {\em Statistics and computing\/}~{\em 27}, 1293--1305.

\bibitem[\protect\citeauthoryear{Hinkley}{Hinkley}{1970}]{hinkley1970inference}
Hinkley, D.~V. (1970).
\newblock Inference about the change-point in a sequence of random variables.
\newblock {\em Biometrika\/}~{\em 57\/}(1), 1 -- 17.

\bibitem[\protect\citeauthoryear{Jansen, Tegomoh, Lange, Showalter, Figliomeni,
  Abdalhamid, Iwen, Fauver, Buss, and Donahue}{Jansen
  et~al.}{2021}]{jansen_2021_investigation}
Jansen, L., B.~Tegomoh, K.~Lange, K.~Showalter, J.~Figliomeni, B.~Abdalhamid,
  P.~C. Iwen, J.~Fauver, B.~Buss, and M.~Donahue (2021, 12).
\newblock Investigation of a {SARS-CoV}-2 {B}.1.1.529 ({O}micron) {V}ariant
  {C}luster — {N}ebraska, {N}ovember–{D}ecember 2021.
\newblock {\em MMWR. Morbidity and Mortality Weekly Report\/}~{\em 70},
  1782--1784.

\bibitem[\protect\citeauthoryear{Killick, Fearnhead, and Eckley}{Killick
  et~al.}{2012}]{killick2012optimal}
Killick, R., P.~Fearnhead, and I.~A. Eckley (2012).
\newblock Optimal detection of changepoints with a linear computational cost.
\newblock {\em Journal of the American Statistical Association\/}~{\em
  107\/}(500), 1590--1598.

\bibitem[\protect\citeauthoryear{Kurt, Y{\i}lmaz, and Wang}{Kurt
  et~al.}{2020}]{kurt2020real}
Kurt, M.~N., Y.~Y{\i}lmaz, and X.~Wang (2020).
\newblock Real-time nonparametric anomaly detection in high-dimensional
  settings.
\newblock {\em IEEE transactions on pattern analysis and machine
  intelligence\/}~{\em 43\/}(7), 2463--2479.

\bibitem[\protect\citeauthoryear{Li and Gu}{Li and Gu}{2021}]{li2021robust}
Li, H. and M.~Gu (2021).
\newblock Robust estimation of {SARS-CoV}-2 epidemic in {US} counties.
\newblock {\em Scientific reports\/}~{\em 11\/}(1), 11841.

\bibitem[\protect\citeauthoryear{Li, Ma, Shende, Castaneda, Chakladar, Tsai,
  Apostol, Honda, Xu, Wong, Zhang, Lee, Gnanasekar, Honda, Kuo, Yu, Chang,
  Rajasekaran, and Ongkeko}{Li et~al.}{2020}]{li2020using}
Li, W.~T., J.~Ma, N.~Shende, G.~Castaneda, J.~Chakladar, J.~C. Tsai,
  L.~Apostol, C.~O. Honda, J.~Xu, L.~M. Wong, T.~Zhang, A.~Lee, A.~Gnanasekar,
  T.~K. Honda, S.~Z. Kuo, M.~A. Yu, E.~Y. Chang, M.~R. Rajasekaran, and W.~M.
  Ongkeko (2020).
\newblock Using machine learning of clinical data to diagnose {COVID}-19: a
  systematic review and meta-analysis.
\newblock {\em BMC medical informatics and decision making\/}~{\em 20\/}(1),
  1--13.

\bibitem[\protect\citeauthoryear{Matteson and James}{Matteson and
  James}{2014}]{matteson2014nonparametric}
Matteson, D.~S. and N.~A. James (2014).
\newblock A nonparametric approach for multiple change point analysis of
  multivariate data.
\newblock {\em Journal of the American Statistical Association\/}~{\em
  109\/}(505), 334--345.

\bibitem[\protect\citeauthoryear{Monaghan, Larkin, Chaudhuri, Han, Jiao,
  Bermudez, Weinhandl, Dahne-Steuber, Belmonte, Neri, Kotanko, Kooman, Hymes,
  Kossmann, Usvyat, and Maddux}{Monaghan et~al.}{2021}]{monaghan2021machine}
Monaghan, C.~K., J.~W. Larkin, S.~Chaudhuri, H.~Han, Y.~Jiao, K.~M. Bermudez,
  E.~D. Weinhandl, I.~A. Dahne-Steuber, K.~Belmonte, L.~Neri, P.~Kotanko, J.~P.
  Kooman, J.~L. Hymes, R.~J. Kossmann, L.~A. Usvyat, and F.~W. Maddux (2021).
\newblock Machine learning for prediction of patients on hemodialysis with an
  undetected {SARS-CoV}-2 infection.
\newblock {\em Kidney360\/}~{\em 2\/}(3), 456.

\bibitem[\protect\citeauthoryear{Ni, M{\"u}ller, and Ji}{Ni
  et~al.}{2020}]{ni2020bayesian}
Ni, Y., P.~M{\"u}ller, and Y.~Ji (2020).
\newblock Bayesian double feature allocation for phenotyping with electronic
  health records.
\newblock {\em Journal of the American Statistical Association\/}~{\em
  115\/}(532), 1620--1634.

\bibitem[\protect\citeauthoryear{Page}{Page}{1954}]{page1954continuous}
Page, E.~S. (1954).
\newblock Continuous inspection schemes.
\newblock {\em Biometrika\/}~{\em 41\/}(1/2), 100--115.

\bibitem[\protect\citeauthoryear{Prado and West}{Prado and
  West}{2010}]{prado2010time}
Prado, R. and M.~West (2010).
\newblock {\em Time series: modeling, computation, and inference}.
\newblock Chapman and Hall/CRC.

\bibitem[\protect\citeauthoryear{Romano, Rigaill, Runge, and Fearnhead}{Romano
  et~al.}{2022}]{romano2022detecting}
Romano, G., G.~Rigaill, V.~Runge, and P.~Fearnhead (2022).
\newblock Detecting abrupt changes in the presence of local fluctuations and
  autocorrelated noise.
\newblock {\em Journal of the American Statistical Association\/}~{\em
  117\/}(540), 2147--2162.

\bibitem[\protect\citeauthoryear{Saat{\c{c}}i, Turner, and
  Rasmussen}{Saat{\c{c}}i et~al.}{2010}]{saatcci2010gaussian}
Saat{\c{c}}i, Y., R.~D. Turner, and C.~E. Rasmussen (2010).
\newblock Gaussian process change point models.
\newblock In {\em Proceedings of the 27th International Conference on Machine
  Learning (ICML-10)}, pp.\  927--934.

\bibitem[\protect\citeauthoryear{Sch{\"o}lkopf and Smola}{Sch{\"o}lkopf and
  Smola}{2018}]{scholkopf2002learning}
Sch{\"o}lkopf, B. and A.~J. Smola (2018).
\newblock {\em {{Learning with {K}ernels: {S}upport Vector Machines,
  Regularization, Optimization, and Beyond}}}.
\newblock The MIT Press.

\bibitem[\protect\citeauthoryear{Shojaie and Fox}{Shojaie and
  Fox}{2022}]{shojaie2022granger}
Shojaie, A. and E.~B. Fox (2022).
\newblock Granger causality: A review and recent advances.
\newblock {\em Annual Review of Statistics and Its Application\/}~{\em 9},
  289--319.

\bibitem[\protect\citeauthoryear{Song, Lee, Kim, Jeong, Kim, Kim, Yoo, Lee,
  Lee, Lee, Kim, Rhee, Kim, and Park}{Song et~al.}{2022}]{song_2022_serial}
Song, J.~S., J.~Lee, M.~Kim, H.~S. Jeong, M.~S. Kim, S.~G. Kim, H.~N. Yoo,
  J.~J. Lee, H.~Y. Lee, S.-E. Lee, E.~J. Kim, J.~E. Rhee, I.~H. Kim, and Y.-J.
  Park (2022).
\newblock Serial {I}ntervals and {H}ousehold {T}ransmission of {SARS-CoV}-2
  {O}micron {V}ariant, {S}outh {K}orea, 2021.
\newblock {\em Emerging Infectious Diseases\/}~{\em 28}, 756--759.

\bibitem[\protect\citeauthoryear{Stein}{Stein}{1999}]{stein1999interpolation}
Stein, M.~L. (1999).
\newblock {\em Interpolation of spatial data: some theory for kriging}.
\newblock Springer Science \& Business Media.

\bibitem[\protect\citeauthoryear{Stroud, M{\"u}ller, and Sans{\'o}}{Stroud
  et~al.}{2001}]{stroud2001dynamic}
Stroud, J.~R., P.~M{\"u}ller, and B.~Sans{\'o} (2001).
\newblock Dynamic models for spatiotemporal data.
\newblock {\em Journal of the Royal Statistical Society: Series B (Statistical
  Methodology)\/}~{\em 63\/}(4), 673--689.

\bibitem[\protect\citeauthoryear{van~den Burg and Williams}{van~den Burg and
  Williams}{2020}]{van2020evaluation}
van~den Burg, G.~J. and C.~K. Williams (2020).
\newblock An evaluation of change point detection algorithms.
\newblock {\em arXiv preprint arXiv:2003.06222\/}.

\bibitem[\protect\citeauthoryear{West and Harrison}{West and
  Harrison}{1997}]{West1997}
West, M. and P.~J. Harrison (1997).
\newblock {\em Bayesian Forecasting $\&$ Dynamic Models\/} (2nd ed.).
\newblock Springer Verlag.

\bibitem[\protect\citeauthoryear{Whittle}{Whittle}{1954}]{whittle1954stationary}
Whittle, P. (1954).
\newblock On stationary processes in the plane.
\newblock {\em Biometrika\/}~{\em 41\/}(3/4), 434--449.

\bibitem[\protect\citeauthoryear{Zhang, Siegmund, Ji, and Li}{Zhang
  et~al.}{2010}]{zhang2010detecting}
Zhang, N.~R., D.~O. Siegmund, H.~Ji, and J.~Z. Li (2010).
\newblock Detecting simultaneous changepoints in multiple sequences.
\newblock {\em Biometrika\/}~{\em 97\/}(3), 631--645.

\bibitem[\protect\citeauthoryear{Zhang, Griffin, and Matteson}{Zhang
  et~al.}{2023}]{zhang2023modeling}
Zhang, W., M.~Griffin, and D.~S. Matteson (2023).
\newblock Modeling a nonlinear biophysical trend followed by long-memory
  equilibrium with unknown change point.
\newblock {\em The Annals of Applied Statistics\/}~{\em 17\/}(1), 860--880.

\bibitem[\protect\citeauthoryear{Zoabi, Deri-Rozov, and Shomron}{Zoabi
  et~al.}{2021}]{zoabi2021machine}
Zoabi, Y., S.~Deri-Rozov, and N.~Shomron (2021).
\newblock Machine learning-based prediction of {COVID}-19 diagnosis based on
  symptoms.
\newblock {\em npj digital medicine\/}~{\em 4\/}(1), 3.

\end{thebibliography}

\end{document}